\documentclass{vgtc}                          

\graphicspath{{}{pictures/}{images/}{./}} 

\usepackage{times}                     

\usepackage{tabu}                      
\usepackage{booktabs}                  
\usepackage{lipsum}                    
\usepackage{mwe}                       
\usepackage{tabularx}
\usepackage[normalem]{ulem}
\usepackage{xcolor}
\usepackage{soul}

\sethlcolor{yellow}
\usepackage{amssymb}

\usepackage{textcomp}
\usepackage{amsmath}
\usepackage{multirow}
\usepackage[most]{tcolorbox}
\usepackage{xcolor}
\usepackage{enumitem}

\usepackage{mathptmx}                  

\onlineid{0}

\vgtccategory{Research}

\vgtcinsertpkg

\title{FootprintRAG: Visual Analytics for Evidence Context Refinement in RAG-based Scientific Literature Exploration}

\providecommand{\affmark}[1]{}
\renewcommand{\affmark}[1]{\textsuperscript{\normalfont\scriptsize #1}}
\newcommand{\inlineicon}[1]{\raisebox{-0.18em}{\includegraphics[height=0.9em]{#1}}}
\author{
\begin{tabular}{c}
\begin{tabular*}{0.98\linewidth}{@{\extracolsep{\fill}}cccccc@{}}
Xingyu Liu\affmark{1,2} &
Yu Dong\affmark{1} &
Qizhen Yu\affmark{1,2} &
Shiyu Cheng\affmark{1} &
Zhe Wang\affmark{1,2} &
Guan Li\affmark{1,2}
\end{tabular*}
\\[0.45em]
\begin{tabular*}{0.68\linewidth}{@{\extracolsep{\fill}}cccc@{}}
Guihua Shan\affmark{1,2,3} &
Dong Tian\affmark{1,2} &
Christy Jie Liang\affmark{4} &
Quang Vinh Nguyen\affmark{5}
\end{tabular*}
\end{tabular}
}

\affiliation{
\scriptsize
\begin{minipage}{0.98\linewidth}
\centering
\affmark{1} Computer Network Information Center, Chinese Academy of Sciences \quad
\affmark{2} University of Chinese Academy of Sciences\\
\affmark{3} Hangzhou Institute for Advanced Study, UCAS\\
\affmark{4} University of Technology Sydney \quad
\affmark{5} Western Sydney University
\end{minipage}
}

\teaser{
  \centering
  \includegraphics[width=\linewidth]{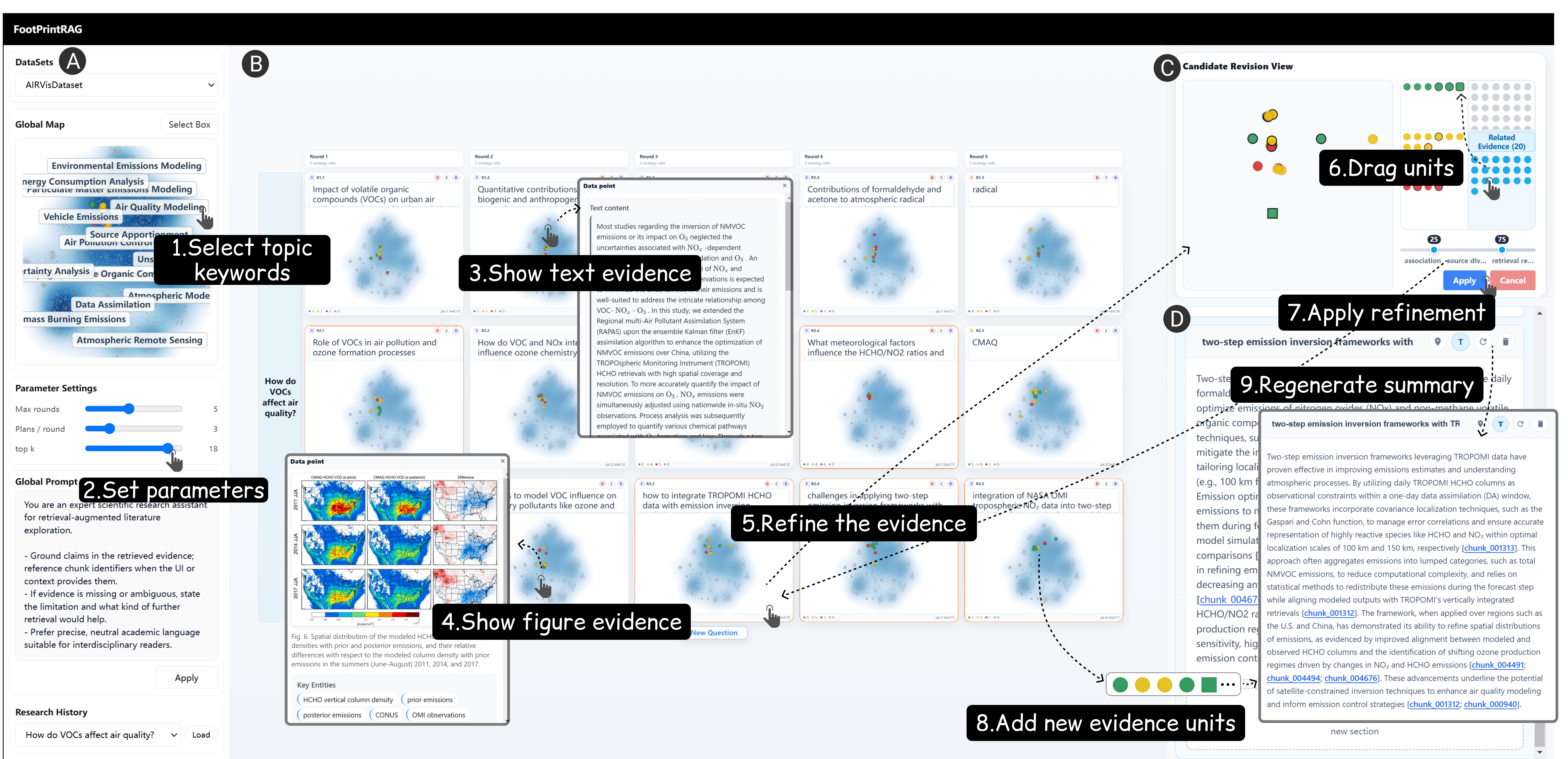}
  \caption{Overview of \textit{FootprintRAG}, illustrated through Case 1. (A) Control Panel supports corpus selection, global embedding overview, parameter setting, and global prompting. (B) RAG-Iteration Matrix View compares multi-round retrieval strategies. (C) Evidence Space Revision View supports evidence-state revision with reranked units, local candidate pool units, and ERS-ranked supplementary candidates. (D) Context Summarization View generates evidence-grounded summaries. The numbered annotations trace Case 1: selecting VOC-related topics, setting parameters, inspecting text and figure evidence, refining and adding evidence units, applying the revised evidence state, and regenerating the summary.}
  \label{fig:teaser}
  
}

\abstract{

Retrieval-Augmented Generation (RAG) is increasingly used to ground large language model (LLM) outputs in scientific literature. However, in open-ended literature exploration, the evidence context used for generation is often produced through hidden retrieval, reranking, assessment, and filtering steps. Users may receive retrieval summaries without knowing how the system constructed the evidence context, which evidence units were retained or discarded, or whether potentially useful evidence was excluded before synthesis. We present \textit{FootprintRAG}, an LLM-agent-powered visual analytics system for evidence context refinement in RAG-based scientific literature exploration. The core idea is to treat the RAG evidence context as an explicit, inspectable, and revisable analytical object before generation. \textit{FootprintRAG} parses scientific literature into text and figure evidence units, expands an initial query into parallel query variants, retrieves and assesses evidence across iterative rounds, and surfaces ERS-ranked supplementary candidates from the corpus-level evidence space. Through coordinated views, the system connects retrieval trajectories, evidence-state revision, and provenance-aware summary generation into a user-steerable workflow. We evaluate \textit{FootprintRAG} through two case studies, a user study, and a workflow-level comparison with representative RAG systems. The results show that \textit{FootprintRAG} helps users compare retrieval directions, revise candidate evidence, recover potentially overlooked evidence, and trace generated summaries back to supporting evidence units. \textit{FootprintRAG} is available at \href{https://github.com/meteorshowering/FootprintRAGVA.git}{GitHub}.

} 


\begin{document}


\firstsection{Introduction}

\maketitle


Recent RAG-based systems have advanced scientific document processing, retrieval, reranking, multimodal evidence handling, and citation-backed generation. However, these advances mainly optimize the retrieval--generation pipeline \cite{asai2026synthesizing}. In open-ended literature exploration, users still have limited access to the pre-generation evidence context: they often cannot inspect how an initial question was expanded, which retrieval directions were explored, which evidence units were retained or discarded, or whether useful evidence was filtered out before synthesis.

This limitation makes it difficult to use RAG-based systems directly as exploratory tools \cite{brown2025systematic}. Our target users are researchers conducting open-ended synthesis over a pre-curated scientific corpus. They need to compare query variants, inspect mixed-quality candidates, remove redundant evidence, recover overlooked findings, and decide which evidence should support the synthesis. The evidence context is therefore part of the analytical process because its coverage and source composition shape the final account.

In this paper, we define \textit{evidence context refinement} to describe this in-process workflow: retrieved evidence is inspected, assessed, supplemented, revised, and confirmed before it is used for synthesis. Treating this workflow as a visual analytics problem changes the role of users in RAG-based literature exploration. Instead of correcting generated text after the fact, users can intervene in the evidence context itself. This requires making query variants, retrieval results, evidence states, supplementary candidates, user revisions, and provenance relations explicit and revisable.

We present \textit{FootprintRAG}, a visual analytics system that combines agent-assisted evidence proposal with user interaction, allowing researchers to inspect retrieval processes, revise evidence states, recover potentially useful evidence, and generate evidence-grounded summaries from confirmed contexts. The system supports coverage-aware research synthesis by making evidence construction inspectable and revisable.

The main contributions of this work are:

\begin{itemize}

    \item We propose an LLM-agent-assisted evidence context refinement workflow that supports parallel query expansion, candidate assessment, local candidate recovery, corpus-level supplementary candidate selection, and user-confirmed synthesis.

    \item We develop \textit{FootprintRAG}, a visual analytics system that externalizes multi-round retrieval processes, supports evidence-state revision, and generates evidence-grounded summaries with provenance links to text and figure evidence units.

    \item We evaluate \textit{FootprintRAG} through two case studies, a user study, and a workflow-level comparison with representative RAG systems and visual analytics tools.
\end{itemize}

\section{Related Work}

\subsection{Visual Analytics for Scientific Literature Exploration}

Visual analytics for scientific literature exploration helps researchers extract structured knowledge from large corpora and identify key contributions, topics, and relationships. Existing approaches use spatial metaphors~\cite{liu2025city}, literature networks~\cite{yang2025litforager}, topic models~\cite{tian2023litvis}, and citation-based recommendations~\cite{beck2024puresuggest,behera2023visual}. \textit{LitVis} connects topic analysis with citation relationships to reveal research development, while \textit{PUREsuggest} combines citation networks with keyword-controlled rankings. \textit{CiteSee}~\cite{chang2023citesee} contextualizes citations using readers' prior activities, and \textit{PaperWeaver}~\cite{lee2024paperweaver} explains connections between recommended and collected papers. \textit{DocFlow}~\cite{qiu2022docflow} supports question-driven retrieval and categorization for systematic reviews, while \textit{VITALITY}~\cite{narechania2021vitality} combines document embeddings with coordinated views for corpus exploration. These approaches support field overviews and document-level relationships, but are less focused on selecting document-internal evidence for generation.

Fine-grained exploration methods support keyphrase-guided information seeking~\cite{tu2022phrasemap}, text-cluster analysis~\cite{peng2025textlens}, and exploration of scientific figures and tables~\cite{chen2021vis30k}. \textit{Relatedly}~\cite{palani2023relatedly} supports exploration of related-work paragraphs through diversity-aware ranking and highlighting, while \textit{Scim}~\cite{fok2023scim} highlights salient passages with reader-controlled density. \textit{Threddy}~\cite{kang2022threddy} organizes extracted passages and supporting papers into research threads. At the collection level, \textit{LitForager}~\cite{yang2025litforager} supports spatial literature organization through multimodal interaction. Recent systems also address LLM-related artifacts: \textit{Graphologue}~\cite{jiang2023graphologue} transforms LLM responses into interactive diagrams, while \textit{KEditVis}~\cite{chen2026keditvis} supports model-layer selection and comparison of knowledge-editing outcomes. \textit{ConceptEVA}~\cite{zhang2023concepteva} supports concept-driven summary customization, and \textit{SurveyAgent}~\cite{wang2024surveyagent} combines paper management, recommendation, and conversational question answering.

These studies demonstrate the value of visual analytics for corpus-level and fine-grained literature exploration, but most focus on browsing, retrieval, or reading support. \textit{FootprintRAG} instead focuses on how textual and visual evidence units are selected, revised, and organized as context for downstream generation.

\subsection{RAG-based Scientific Literature Exploration}

RAG has become an important paradigm for mitigating hallucinations in large language models and improving knowledge-intensive generation by incorporating external knowledge into the LLM context. This mechanism is well suited to scientific literature exploration because it can retrieve from large external knowledge bases and ground generated content in verifiable evidence~\cite{han2025fine,hwang2025retrieval}. For example, \textit{ResearchAgent}~\cite{baek2025researchagent} adopts iterative retrieval and feedback to support research idea generation, while \textit{OpenScholar}~\cite{asai2026synthesizing} retrieves passages from 45 million open-access scientific papers to generate citation-backed answers. Related scientific agents extend this direction: \textit{SciAgents}~\cite{ghafarollahi2025sciagents} combines knowledge graphs and multi-agent reasoning to develop materials hypotheses, and \textit{ChemCrow}~\cite{bran2023chemcrow} integrates chemistry tools for domain-specific tasks.

Recent RAG-based literature exploration systems have advanced this paradigm along several technical directions. \textit{HiPerRAG}~\cite{gokdemir2025hiperrag} improves retrieval throughput and accuracy for processing scientific literature, while \textit{SciRAG}~\cite{ding2026scirag} introduces an adaptive, citation-aware, and outline-guided RAG framework for scientific literature review generation. Other approaches extend RAG to richer data and knowledge structures. \textit{VisRAG}~\cite{yu2025visrag} retrieves document images and generates answers from visual inputs, preserving information lost in text-only parsing. Graph-based RAG methods~\cite{peng2025graph} use graph structures and relationships to organize retrieval and generation. Beyond static document modalities, \textit{NotebookRAG}~\cite{shan2026notebookrag} treats executable code cells in exploratory data analysis notebooks as retrievable units and uses multi-notebook retrieval with agents to support automatic notebook generation.

These systems advance retrieval efficiency, citation grounding, multimodal processing, graph reasoning, and task-specific generation, but mainly optimize internal retrieval--generation pipelines or specific scientific tasks. They provide limited support for researchers to inspect, compare, revise, and organize candidate evidence before it becomes generation context. \textit{FootprintRAG} addresses this gap by making evidence selection across retrieval directions inspectable and revisable before synthesis.

\subsection{Visual Analytics for RAG-based Workflows}

Recent surveys summarize the core components of RAG, including retrieval mechanisms, intermediate fusion strategies, generation modules, and performance across application scenarios~\cite{brown2025systematic,oche2025systematic,sharma2025retrieval}. RAG has evolved from vector retrieval and simple generation pipelines~\cite{gao2023retrieval,brown2025systematic} to richer design patterns, including graph-structured augmentation~\cite{peng2025graph}, hybrid retrieval~\cite{kalra2024hypa,lee2025hybgrag,wang2024searching}, robustness optimization~\cite{yan2024corrective}, and knowledge-oriented applications~\cite{cheng2025survey}. Related work also addresses LLM-powered genome visualization~\cite{zhang2025auragenome}, scientific visualization generation and caption alignment~\cite{lu2026agentic}, and structure-aware retrieval for visualization pipelines~\cite{zhao2026toward}. These studies improve internal mechanisms or domain-specific generation. However, the steps between retrieval, ranking, context construction, and generation are often not exposed as user-revisable evidence states.

Visual analytics makes intermediate computational processes inspectable. For example, \textit{VizOPTICS}~\cite{wu2023vizoptics} supports interactive inspection of OPTICS cluster formation and refinement of clustering results. \textit{UcVE}~\cite{yu2023user} supports in-process comparison by saving and tracking exploration states and comparing visualization units across historical and current results. These approaches illustrate how visual interfaces support inspection of computational processes and comparison across exploration states.

Within RAG workflows, recent visual analytics tools support the inspection and refinement of retrieval and generation. \textit{RAGExplorer}~\cite{tian2026ragexplorervisualanalyticscomparative} compares RAG configurations to support diagnosis and optimization. \textit{RAGViz}~\cite{wang2024ragviz} visualizes retrieved documents and their influence on generated responses, while \textit{RAGTrace}~\cite{cheng2025ragtrace} supports inspection and refinement of retrieval--generation dynamics. \textit{XGraphRAG}~\cite{wang2025xgraphrag} traces retrieved information through graph-based RAG pipelines. Complementary tools expose agent behavior and model reasoning: \textit{AgentLens}~\cite{lu2024agentlens} supports temporal exploration and cause tracing of agent behaviors, \textit{ReasonGraph}~\cite{li2025reasongraph} visualizes reasoning methods and inference paths, and \textit{Interactive Reasoning}~\cite{pang2026interactive} enables review and modification of chain-of-thought outputs. These systems demonstrate the value of making intermediate processes inspectable, while targeting different analytical objects from the evidence contexts used in literature synthesis.

The RAG-focused tools discussed above emphasize configuration diagnosis, failure analysis, and pipeline inspection, often from a developer perspective. \textit{FootprintRAG} instead targets researchers conducting scientific literature exploration and focuses on interactive control of evidence context construction for synthesis, including comparison across retrieval directions, candidate evidence inspection, evidence filtering, and provenance-aware confirmation.

\section{Formative Study}
\label{sec:formative-study}

We conducted a formative study to investigate how researchers use search tools and LLMs for scientific literature exploration, and where visual analytics could support RAG-based workflows.

\subsection{Experts and Procedure}

We recruited 8 experts (\(E_1\)--\(E_8\)) with substantial experience in scientific literature search and synthesis. They included Ph.D. candidates, postdoctoral researchers, and early-career researchers across multiple research domains. All had conducted literature reviews for research projects, papers, proposals, or related work sections. They regularly used academic search engines, citation-based exploration tools, reference managers, and LLM-based tools for summarization, topic exploration, or research ideation; six had experience with document question answering or RAG-like workflows.

Each interview lasted approximately 45 minutes and followed a semi-structured format. We first asked experts to describe a recent open-ended literature exploration task, including how they formulated search queries, inspected candidate papers or passages, and judged topic coverage. We then discussed their experience with literature search tools and LLM-based assistants, focusing on candidate evidence inspection, evidence quality judgment, and summary verification. Finally, we discussed a general RAG-based literature exploration process and asked which intermediate results should be inspectable, which automatic decisions should remain revisable, and what evidence should be confirmed before accepting a synthesis.

We analyzed the interview notes through iterative qualitative coding. We marked recurring observations related to query formulation, candidate evidence judgment, tool limitations, missing or redundant evidence, and summary verification. We then consolidated these observations into the findings and design goals presented below.

\subsection{Findings}

We summarized four findings that characterize evidence context refinement in RAG-based literature exploration.

\textbf{F1: Literature exploration requires parallel query variants.}
Experts described open-ended literature exploration as divergent: they generate multiple query formulations to examine aspects such as data sources, mechanisms, methods, tasks, visual encodings, or evaluation settings. These variants help compare evidence coverage and decide which directions to refine. However, LLM-based tools often hide internal query rewriting, making it difficult to judge whether reasonable directions were explored or prematurely narrowed.

\textbf{F2: Retrieved candidates contain mixed evidence quality.}
Experts emphasized that semantic relevance does not necessarily imply evidence utility. Retrieved candidates may contain matching terms but only provide background, redundancy, or weakly related discussion, while lower-ranked candidates may contain specific findings, figure descriptions, or methodological details useful for synthesis. Experts therefore wanted to inspect candidates, remove weak evidence, and treat LLM assessment as initial triage.

\textbf{F3: Useful evidence may be hidden by automatic truncation.}
Experts noted that top-\(k\) retrieval or reranking may hide useful evidence, especially for interdisciplinary topics or varied terminology. Rather than exposing all discarded candidates, they preferred a limited set worth revisiting, such as candidates related to useful retrieved evidence, less redundant with the current context, or from underrepresented sources.

\textbf{F4: Evidence context should be confirmed before synthesis.}
Experts considered citations useful but insufficient for trusting generated summaries. Post-hoc citations do not clearly reveal which evidence was used, which candidates were excluded, or whether the final context covered the queries. Before generation, experts wanted to inspect and adjust the evidence context, especially in tasks focused on mapping a research landscape or identifying gaps where coverage, source balance, and weak evidence can strongly affect the final synthesis.

\subsection{Design Goals}

Based on these findings, we summarized four design goals.

\textbf{DG1: Support comparison across parallel query variants.}
The system should preserve multiple query variants within an exploration round and allow users to compare the evidence returned by different variants.

\textbf{DG2: Enable evidence revision before synthesis.}
The system should make retrieved candidates and automatic assessments inspectable and revisable, allowing users to remove unsuitable evidence and correct assessment results before generation.

\textbf{DG3: Surface potentially overlooked evidence.}
The system should expose a limited set of additional candidates that may have been missed by automatic ranking. These candidates should support user inspection without being automatically included in the final evidence context.

\textbf{DG4: Support evidence context confirmation and provenance.}
The system should allow users to inspect and adjust the evidence selected for summary generation when coverage or source balance matters, while maintaining links from selected evidence to its original sources.

\section{FootprintRAG Overview}
\label{sec:overview}

\begin{figure*}
    \centering
    \includegraphics[width=\linewidth]{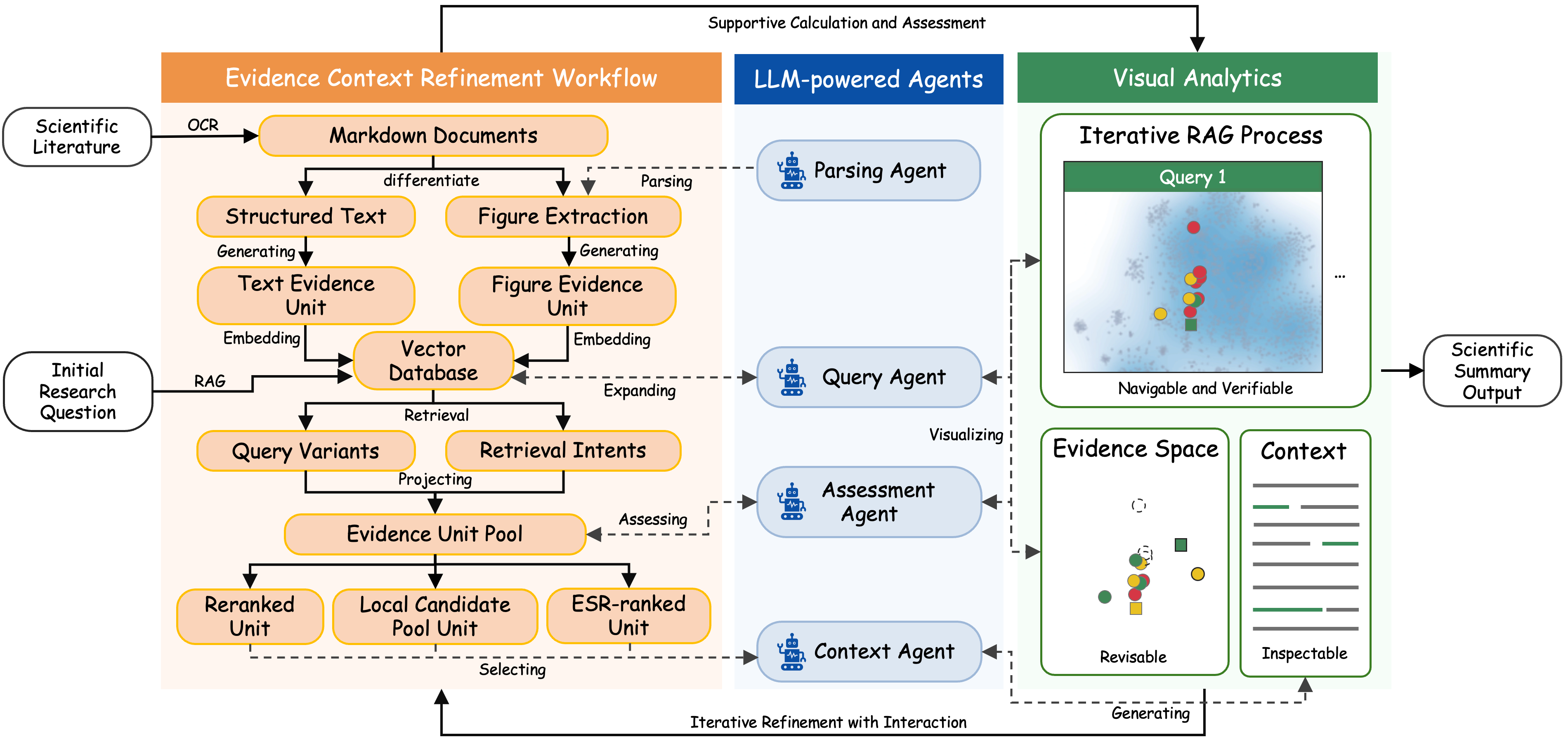}
    \caption{System architecture of \textit{FootprintRAG}. The evidence context refinement workflow converts scientific literature into text and figure evidence units, retrieves and assesses candidate evidence with LLM-powered agents, and exposes the iterative RAG process, evidence space, and context through visual analytics. Users refine candidate evidence before generating an evidence-grounded scientific summary.}
    \label{fig:overview}
    \vspace{-1.2em}
\end{figure*}

Figure~\ref{fig:overview} summarizes the architecture of \textit{FootprintRAG}. The system takes scientific literature and an initial research question as input, and supports evidence-grounded summary generation through three coordinated layers. The evidence context refinement workflow converts papers into retrievable evidence units and organizes the intermediate evidence states produced during iterative RAG exploration. LLM-powered agents assist key steps, including figure evidence construction, query expansion, candidate assessment, and context generation. The visual interfaces expose the iterative RAG process, evidence space, and evolving context so that users can inspect, revise, and confirm the evidence used for synthesis. This architecture separates evidence proposal from evidence use: retrieval and agents prepare candidate evidence, while users determine the final evidence context for summary generation. The framework does not require a specific parser, embedding model, or LLM prompting strategy; these components are replaceable implementation modules. The contribution lies in externalizing the evidence states they produce and making them comparable, revisable, and usable for evidence-grounded synthesis.

\section{Evidence Context Refinement Workflow}
\label{sec:workflow}

\subsection{Evidence Unit Construction}

The workflow begins by converting scientific papers into evidence units. An evidence unit is the basic object that can be retrieved, assessed, revised, and cited. Each unit contains a textual representation for retrieval, an embedding vector for similarity computation, and provenance metadata for tracing it back to the original source.

For textual content, papers are converted into structured text using MinerU~\cite{niu2025mineru2} through OCR and layout-aware parsing. The parsed content is segmented into semantically coherent chunks under configurable token constraints, following document hierarchy and paragraph boundaries when possible. Each text evidence unit preserves metadata such as source paper, section, and page number.

For figure content, \textit{FootprintRAG} does not directly index raw image embeddings. Instead, the Parsing Agent constructs a figure evidence unit by combining the original figure, caption, surrounding textual references, and an LLM-generated description. This text-mediated representation allows figures to participate in the same retrieval and assessment process as text, while preserving the original figure and provenance metadata for user inspection. It is intended to support figure-level evidence discovery rather than precise visual measurement from figure marks.

All evidence units are embedded through their textual representations and stored in ChromaDB. In the current implementation, we use ChromaDB with the \texttt{all-MiniLM-L6-v2} embedding model, which maps each evidence unit into a 384-dimensional vector. The resulting embedding space supports query-based retrieval and evidence-level similarity computation.

\subsection{Query Expansion and Retrieval}

Given an initial query, the Query Agent generates multiple query variants. Each variant represents a search direction derived from the original question and is assigned a retrieval intent. A \textit{semantic answer-seeking query} retrieves evidence units that can answer a rewritten question at the semantic level, while an \textit{exact term-matching query} focuses on a specific term, model, mechanism, or method that emerges during exploration.

For each query variant, the workflow uses vector-based retrieval and reranking as traceable modules within the agent-assisted loop. The retrieval step returns an initial candidate set according to query-to-evidence similarity in the embedding space. A reranking step then retains a smaller set of \textit{reranked units} for assessment and later refinement. The candidate sizes and retrieval rounds are configurable. Across rounds, the Query Agent may continue, rewrite, or stop a direction when retrieved evidence becomes repetitive or weakly supported. We denote a retrieval result produced by one query variant in one iteration as \(u\).

\subsection{Candidate Assessment and Supplementary Candidate Selection}

After retrieval and reranking, the Assessment Agent evaluates the reranked units of each retrieval result. We use three LLM-assessed support levels: \textit{High}, \textit{Medium}, and \textit{Low}. \textit{High} indicates strong support for the query variant, \textit{Medium} indicates partial or supporting evidence, and \textit{Low} indicates weak support.

The workflow maintains two additional evidence sources beyond reranked units. The first is the local candidate pool from the current retrieval result. Let \(I_u\) denote the initial candidate set for retrieval result \(u\), and \(R_u\) denote the reranked units retained after reranking. The local candidate pool is:

\vspace{-0.6em}
\[
L_u = I_u \setminus R_u .
\]
\vspace{-0.6em}

Units in \(L_u\) were retrieved by the current query variant but not retained after reranking. They provide local alternatives close to the current RAG result.

The second source is the corpus-level embedding pool for selecting ERS-ranked supplementary candidates. Let \(E\) denote all evidence units in the current corpus, and \(C\) denote evidence units already included in the current evidence context. To keep this source distinct from the current retrieval result, we define the global supplementary pool as:

\vspace{-0.6em}
\[
G_u = E \setminus (I_u \cup C).
\]
\vspace{-0.6em}

Thus, \(L_u\) provides nearby alternatives within the current retrieval result, while \(G_u\) supports discovery of globally related evidence outside the current initial candidate set.

For each candidate \(e \in G_u\), the workflow computes an Evidence Relation Score:

\vspace{-0.6em}
\[
ERS(e) = \alpha A(e) + \beta S(e) + \gamma Q(e),
\]
\vspace{-0.6em}

where all terms are normalized to \([0,1]\). \(A(e)\) measures association with useful reranked units, \(S(e)\) measures source diversity, and \(Q(e)\) preserves a basic query relevance constraint.

The association term is defined as:

\vspace{-0.6em}
\[
A(e) = \frac{1}{|M_u|} \sum_{m \in M_u} sim(e,m),
\]
\vspace{-0.6em}

where \(M_u\) is the set of \textit{High} and \textit{Medium} evidence units in \(R_u\). This term is evidence-set-centered: it favors candidates related to evidence already assessed as potentially useful, rather than ranking candidates only by query similarity. If \(M_u\) is empty, this term is omitted.

The source diversity term is defined as:

\vspace{-0.6em}
\[
S(e) = \frac{1}{1 + count_C(source(e))},
\]
\vspace{-0.6em}

where \(count_C(source(e))\) counts how many evidence units from the same paper are already included in the current evidence context. This term reduces over-reliance on a small number of sources.

Finally, \(Q(e)\) denotes the normalized retrieval relevance of \(e\) to the query variant, preventing supplementary candidates from drifting away from the current retrieval direction. The weights \(\alpha\), \(\beta\), and \(\gamma\) are configurable.

\subsection{Evidence Context Refinement and Summarization}

The evidence context is the set of evidence units confirmed for summary generation. It may include reranked units, local candidate pool units, and ERS-ranked supplementary candidates. Units with low support, removed units, or unconfirmed candidates are excluded from synthesis. After the evidence context is confirmed, the \textit{Context Agent} generates an evidence-grounded summary from the selected evidence units and their provenance metadata.

\section{FootprintRAG System}
\label{sec:system}

\subsection{Control Panel}

The Control Panel supports corpus management, parameter configuration, and global prompting. Users can select a literature corpus, inspect its global embedding distribution, adjust RAG parameters, edit the global prompt, or load previous exploration histories.

The global embedding overview projects evidence units into a two-dimensional space through t-SNE to summarize the selected corpus. A density heatmap and semantic labels help users identify topical regions before retrieval. Because the projection is global and shared across retrieval states, users can also compare evidence distributions across matrix cells to observe evolution, convergence, or repetition of retrieval directions. The projection is used as an overview rather than an exact semantic distance metric. Configurable parameters include retrieval rounds, query variants per round, and candidates retained for inspection.

The global prompt specifies retrieval and synthesis preferences, such as grounding claims in a specific domain, indicating uncertainty when evidence is insufficient, and using neutral academic language. These instructions guide the agents during query expansion, candidate assessment, and summary generation, while the final evidence context is determined through user refinement.

\subsection{RAG-Iteration Matrix View}

The RAG-Iteration Matrix View is the central view of \textit{FootprintRAG}. It presents RAG exploration as a matrix of progressive retrieval results. Rows represent strategies, and columns represent retrieval rounds. Along each row, a strategy may be rewritten, expanded, specialized, or stopped according to retrieved evidence and agent assessment. Within each column, users can compare strategies generated in the same round and judge whether they cover complementary directions or repeatedly retrieve similar evidence.

\begin{figure}
    \centering
    \includegraphics[width=\linewidth]{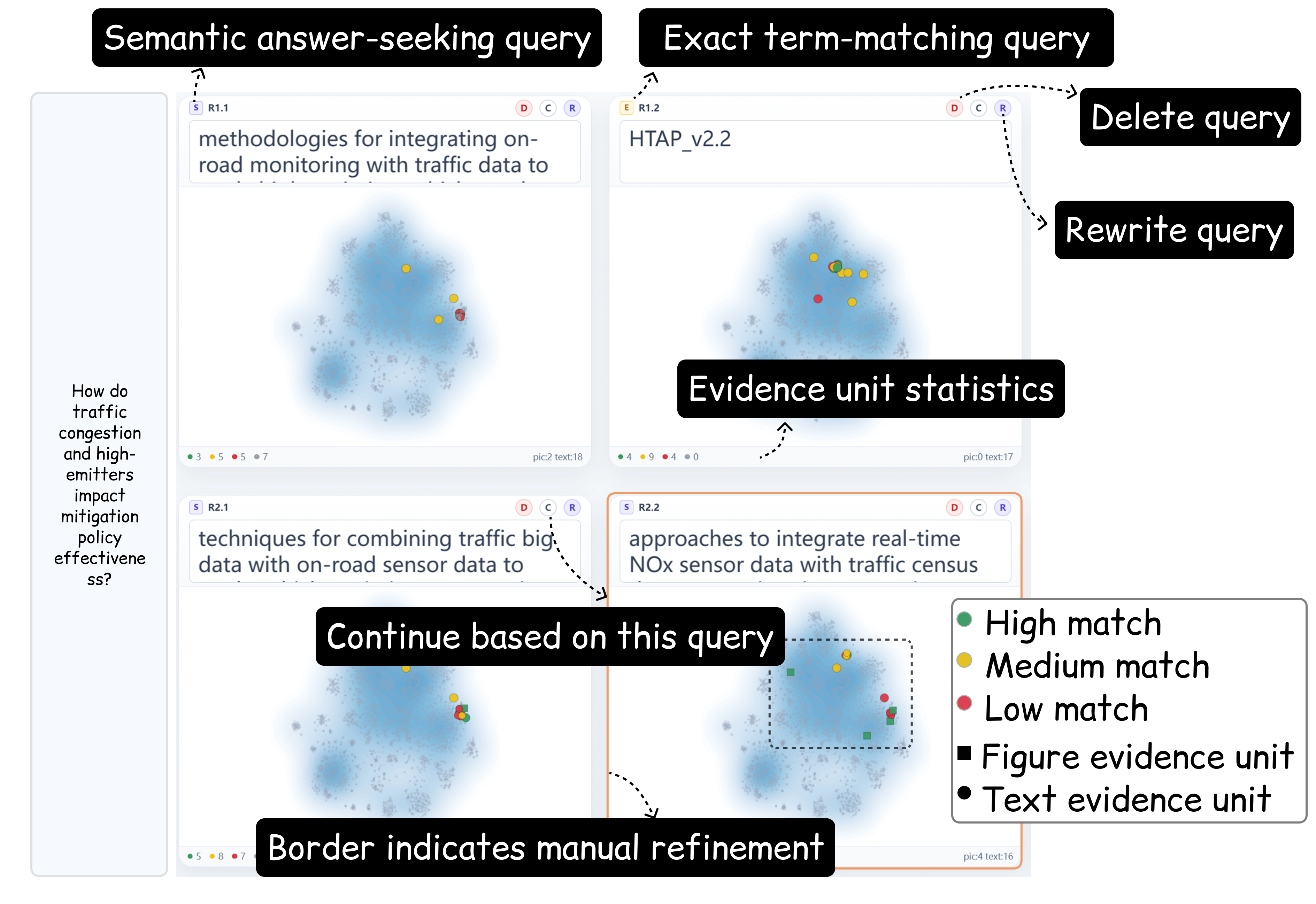}
    \caption{Visual design of the RAG-Iteration Matrix View. Each cell summarizes a retrieval result through intent, operations, evidence-space projection, LLM-assessed support levels, and statistics.}
    \label{fig:matrix}
    \vspace{-1.2em}
\end{figure}

As shown in Fig.~\ref{fig:matrix}, the view distinguishes two retrieval intents. A semantic answer-seeking query, marked by \inlineicon{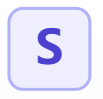}, retrieves evidence units that answer a rewritten question at the semantic level. An exact term-matching query, marked by \inlineicon{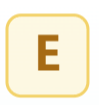}, focuses on a specific term, model, mechanism, or method that emerges during exploration. The agent selects the retrieval intent based on the current query context and retrieved evidence.

Each matrix cell contains a header, a compact evidence space, and statistics. The header shows retrieval intent, round information, and cell-level operations: delete \inlineicon{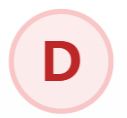}, continue \inlineicon{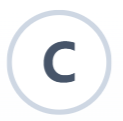}, and rewrite \inlineicon{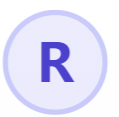}. The evidence space shows retrieved evidence units in the embedding space. Shape encodes modality: circles represent text evidence units, and squares represent figure evidence units. Color encodes LLM-assessed support level: green for \textit{High}, yellow for \textit{Medium}, and red for \textit{Low}. The statistics at the bottom summarize support levels and modalities and can be clicked to highlight corresponding evidence units. Selecting a cell opens its evidence state in the Evidence Space Revision View; once revised and applied, the cell is marked with an orange border to indicate manual refinement.

\subsection{Evidence Space Revision View}

\begin{figure}
    \centering
    \includegraphics[width=\linewidth]{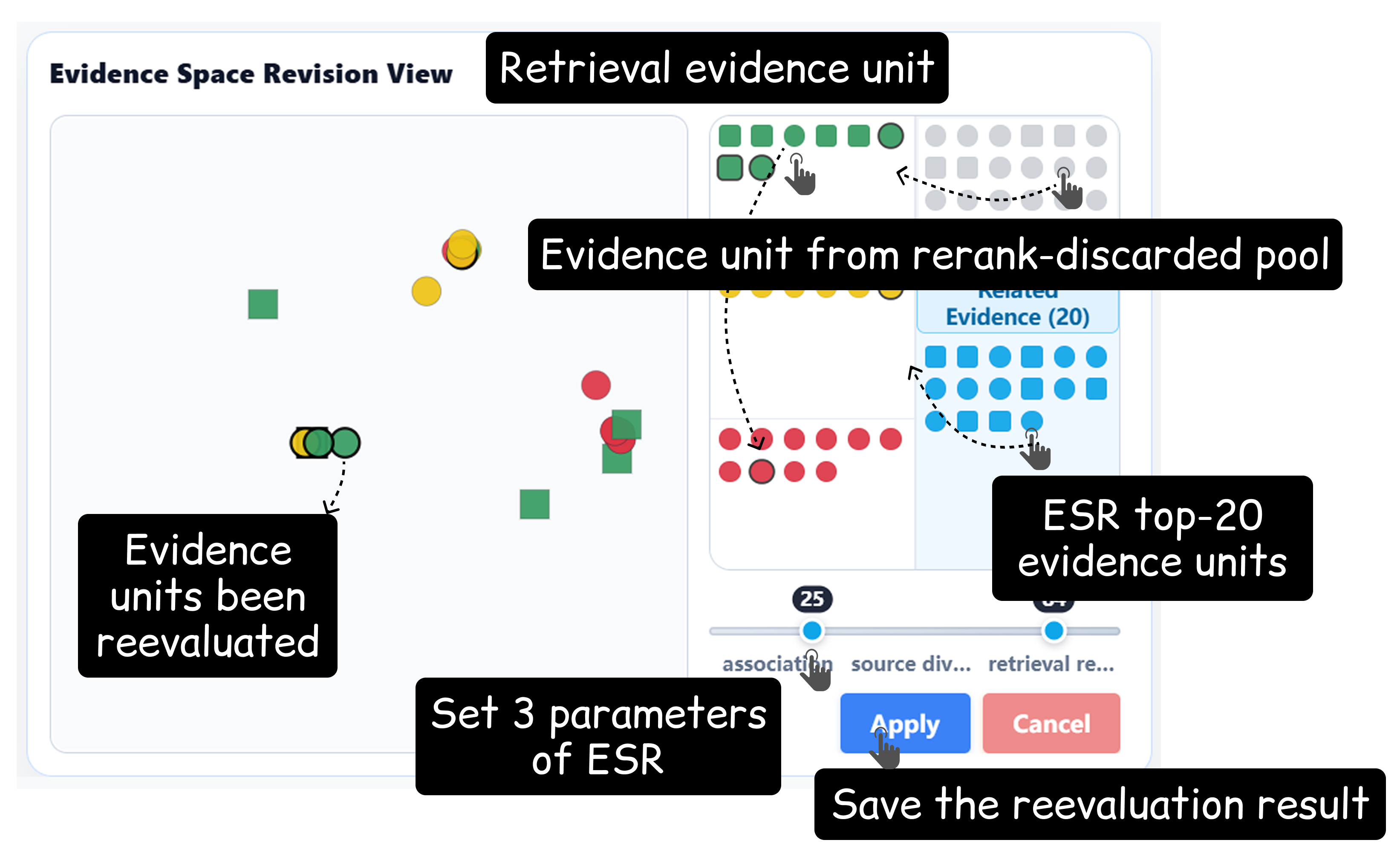}
    \caption{Visual design of the Evidence Space Revision View, which supports evidence-state revision with reranked units, local candidate pool units, and ERS-ranked supplementary candidates.}
    \label{fig:revision-design}
    \vspace{-1.2em}
\end{figure}

The Evidence Space Revision View supports detailed inspection and revision of the evidence state associated with a selected matrix cell. As shown in Fig.~\ref{fig:revision-design}, the left panel presents the local evidence space of the selected retrieval result, using the same modality and support-level encodings as the matrix cells.

The right panel provides a compact revision table. Reranked units are grouped by LLM-assessed support levels. Units in the local candidate pool are shown separately in gray, indicating evidence retrieved by the current query variant but excluded during reranking. ERS-ranked supplementary candidates are shown as \textit{Related Evidence}; they are selected from the corpus-level embedding pool using the Evidence Relation Score. These two sources support different revision purposes: local candidate pool units provide nearby alternatives within the current retrieval result, while ERS-ranked supplementary candidates surface globally related evidence according to the scoring terms defined above.

Users can drag evidence units in either the local evidence space or the table to inspect metadata. They can revise the evidence state by removing unsuitable reranked units, adding useful local candidate pool units, or adding ERS-ranked supplementary candidates. The view also lets users adjust the weights of the three Evidence Relation Score terms: association, source diversity, and retrieval relevance. After interaction, users apply the updated evidence state back to the selected matrix cell.

\subsection{Context Summarization View}

The Context Summarization View supports section-level evidence organization and evidence-grounded summary generation. Users can create editable sections and drag evidence units from the RAG-Iteration Matrix View into each section as generation context. With support from the \textit{Context Agent}, the view generates editable section titles and body text based on the evidence units assigned to each section. Generated content includes citations linked to supporting evidence units. Users can click a citation link to inspect metadata and locate the corresponding evidence unit in the RAG-Iteration Matrix View. If a section is weakly supported or missing important evidence, users can return to the matrix or revision view, update the evidence state, add more evidence units, and regenerate the section. This view connects evidence organization, summary generation, and provenance inspection in a single refinement loop.

\section{Case Study}

\subsection{Case 1: Refining Evidence Pathways from VOC Mechanisms to Air Quality Modeling}

We applied \textit{FootprintRAG} to an atmospheric science corpus consisting of 48 papers. The exploration started with the question: \textit{``How do VOCs affect air quality?''} This question is intentionally broad: a conventional RAG response may mix chemical mechanisms, ozone sensitivity, emission sources, observational indicators, and modeling methods into one summary. We used this case to examine whether \textit{FootprintRAG} can help separate these directions, inspect their supporting evidence, and refine a focused evidence context for synthesis.

As shown in Fig.~\ref{fig:teaser}, we first selected topic keywords from the global map and adjusted retrieval parameters. The RAG-Iteration Matrix then organized the exploration into multiple strategies and rounds. By comparing rows and columns, we observed three major evidence pathways: chemical mechanisms of VOCs and secondary aerosol formation, ozone sensitivity through VOC--NOx interactions and HCHO/NO\(_2\) ratios, and method-oriented evidence linking satellite-constrained emission inversion to regional air-quality model bias correction.

This comparison helped us reinterpret the question: the first two pathways explained VOC impacts, while the third showed how these impacts could be quantified for air-quality modeling. We therefore focused on the method-oriented pathway, where the query evolved toward integrating TROPOMI HCHO data with emission inversion for VOC-related modeling.

We then inspected the corresponding evidence units. Text evidence revealed how satellite observations and inversion methods were used to estimate VOC-related emissions, while figure evidence helped us examine spatial distributions and model-related patterns. In the Evidence Space Revision View, we refined the evidence state by removing units that only provided general VOC definitions or broad photochemical descriptions. We also added useful evidence from both the local candidate pool and ERS-ranked supplementary candidates, including evidence related to HCHO concentration distributions, two-step emission inversion frameworks, and regional model bias correction.

After applying the revised evidence state, we regenerated the corresponding summary section. The updated synthesis no longer described VOC impacts solely in terms of general chemical processes. Instead, it articulated a more specific evidence pathway: satellite-observed HCHO provides observational constraints on VOC-related emissions; inversion methods translate these constraints into NMVOC emission estimates; and the resulting emission information can help reduce regional air-quality model bias. This case shows how \textit{FootprintRAG} supports visual evidence pathway comparison, evidence-state revision, supplementary evidence recovery, and provenance-aware synthesis from a broad scientific question.

\subsection{Case 2: Redirecting LLM-assisted Visualization from Benchmark Evaluation to Workflow Reliability}

We applied \textit{FootprintRAG} to a corpus of 10 representative IEEE TVCG papers to explore an open-ended visualization research topic. After inspecting the global map, we selected evidence units related to LLM-assisted visualization and excluded less relevant regions such as immersive analytics. We then started with the question: \textit{``What are the primary goals of visualization assistance in the LLM era, and what pain points is it solving?''} This question is broad because LLM-assisted visualization spans natural language interfaces, visualization evaluation, interpretation of generated outputs, and workflow-level support for debugging and refinement.

\begin{figure*}
    \centering
    \includegraphics[width=\linewidth]{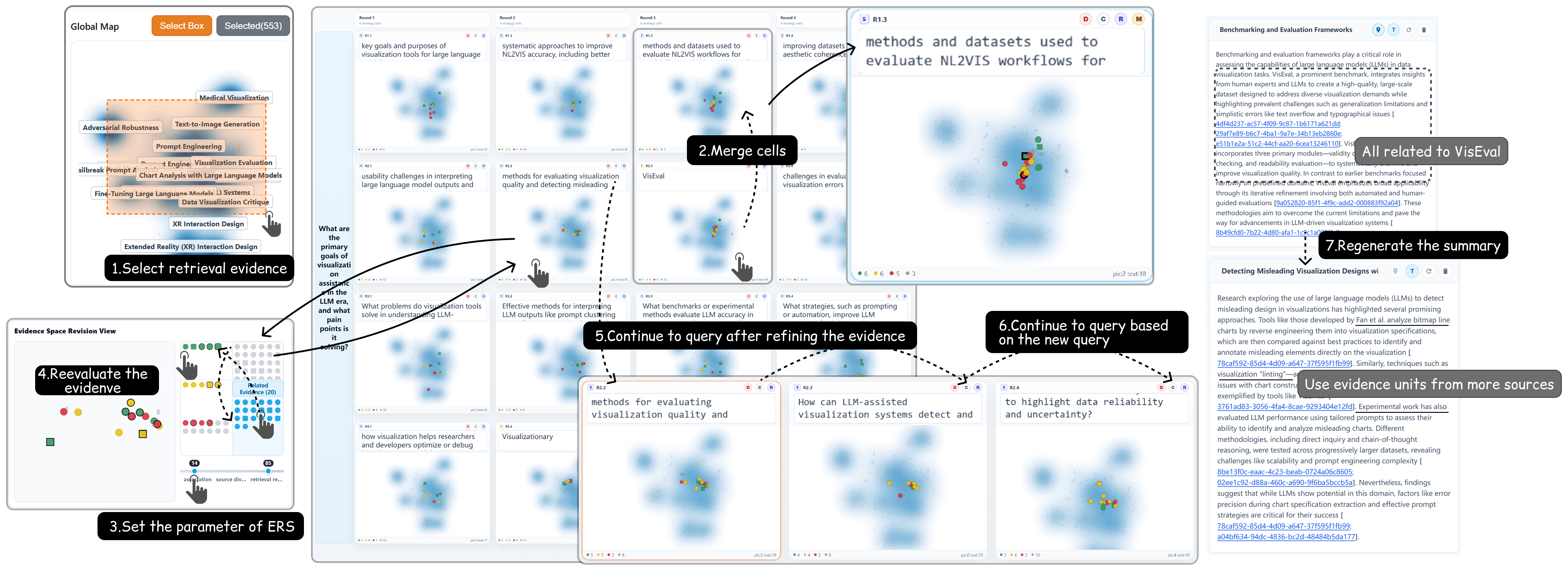}
    \caption{Case 2: The user selects LLM-related evidence units, compares multi-round strategies in the RAG-Iteration Matrix View, identifies convergence around \textit{VisEval}, refines the evidence state with local and ERS-ranked supplementary candidates, continues retrieval from the revised query, and regenerates a broader evidence-grounded summary on workflow reliability.}
    \label{fig:case2}
    \vspace{-1.2em}
\end{figure*}

As shown in Fig.~\ref{fig:case2}, the RAG-Iteration Matrix organized the exploration into multiple strategies across rounds. In the early rounds, the strategies covered different aspects of the topic: goals of LLM-assisted visualization tools, usability and quality challenges, interpretation of LLM-generated outputs, and workflow frameworks for handling execution failures or ambiguous specifications. As the exploration progressed, the matrix made the evolution of each strategy visible. Some strategies became more focused on NL2VIS and benchmark construction, while others moved toward visualization quality assessment, error detection, or workflow-level refinement.

A notable pattern emerged when two strategies converged on similar evidence distributions. One strategy had evolved toward NL2VIS benchmarks, while another focused on evaluating visualization quality in LLM-assisted workflows. Their retrieved evidence units overlapped around \textit{VisEval}, which was relevant to both directions: it provided benchmark-oriented evidence for assessing NL2VIS capabilities and also supported evaluation of LLM-generated visualization code. This convergence helped us identify \textit{VisEval} as a shared evidence anchor across two initially different strategies. However, the convergence also revealed a potential narrowing effect. If we had used this evidence directly for synthesis, the resulting section would likely emphasize benchmark-based evaluation while underrepresenting broader reliability issues in LLM-assisted visualization workflows. We therefore merged related cells for inspection and refined the evidence state in the Evidence Space Revision View. We reduced the number of evidence units focused narrowly on \textit{VisEval} and retained or added evidence related to misleading visualizations, ambiguous specifications, execution failures, and iterative correction.

Using the revised evidence state, we continued retrieval from the updated query direction. The query shifted from \textit{``methods for evaluating visualization quality''} toward a more workflow-oriented question: \textit{``how LLM-assisted visualization systems can detect unreliable outputs and support correction.''} The newly retrieved evidence was no longer dominated by NL2VIS benchmarks, but also included broader discussions on reliability, failure handling, and refinement mechanisms. We then regenerated the corresponding summary section. The revised synthesis no longer framed LLM-assisted visualization primarily as a benchmark evaluation problem. Instead, it described a broader progression: benchmark datasets and evaluation frameworks help measure generated visualization quality, while reliable LLM-assisted visualization also requires mechanisms for resolving ambiguous intent, detecting misleading or failed outputs, and supporting iterative correction. This case shows how \textit{FootprintRAG} helps users identify evidence-pathway convergence and redirect exploration before finalizing a synthesis. The system also supports generating an initial report without manual revision.

\section{User Study}
\label{sec:user-study}


\subsection{Participants and Corpus}

We recruited 10 participants (\(P_1\)--\(P_{10}\)) with experience in scientific literature search and paper reading. They included graduate students and early-career researchers from visualization, human-computer interaction, computer science, bioinformatics, and related interdisciplinary areas. Their research experience ranged from 2 to 6 years (\(M=3.4, SD=1.35\)). All participants had used academic search engines for literature exploration, and 8 had used LLM-based tools for paper summarization or related work writing. None had prior experience with \textit{FootprintRAG}.

We constructed a corpus from the bioinformatics visualization domain, consisting of 16 Markdown-processed papers. The corpus centered on a recent representative paper on LLM-powered genome visualization \cite{zhang2025auragenome} and its cited or closely related references.

\subsection{Apparatus, Task, and Procedure}

The study was conducted in a quiet laboratory setting. Participants used a desktop workstation with a 27-inch 4K monitor and interacted with \textit{FootprintRAG} through a web browser using a standard keyboard and mouse. We collected each participant's final evidence context and generated summary.

Participants were asked to explore the corpus and produce a short evidence-grounded summary. The task prompt was: \textit{``Please investigate how LLM-powered approaches support the generation of genome visualizations. Summarize the main technical challenges, solution strategies, and remaining limitations based on the provided literature corpus.''} This task required participants to synthesize evidence about genome visualization tools, manual configuration barriers, LLM-based visualization generation, and domain-specific limitations across multiple papers.

Each session lasted approximately 70 minutes. We first introduced the study goal and corpus, followed by a 10-minute tutorial on the main views and interactions of \textit{FootprintRAG}. Participants then had 35 minutes to explore the corpus, refine candidate evidence, confirm the evidence context, and generate a final summary. After the task, participants completed a 10-minute questionnaire and joined a 15-minute open-ended interview.

\subsection{Measures}

We collected three types of data: questionnaire ratings, interaction logs, and interview feedback.

First, participants rated five statements on a 5-point Likert scale, where 1 indicated strongly disagree and 5 indicated strongly agree. The questionnaire covered query expansion visibility, evidence usefulness judgment, candidate evidence revision, supplementary candidate usefulness, and trust from evidence context confirmation. The full statements are shown in Fig.~\ref{fig:user-study-rating}.

Second, we logged participants' key interactions during the task, including retrieval-strategy adjustments (regeneration, replanning, and manual query input), evidence-level revisions (removing reranked evidence units, adding local candidates, and adding ERS-ranked supplementary candidates), and summary-level revisions (adding sections, regenerating the whole summary, and rewriting individual sections). These logs were used to analyze how participants distributed their effort across retrieval direction control, evidence-context refinement, and summary revision.

Finally, we conducted an open-ended interview to understand participants' overall experiences, identify useful or difficult aspects of the system, discuss notable evidence revision decisions, and collect suggestions for improvement. We summarized the feedback through thematic grouping.

\subsection{Results}

\begin{figure}
    \centering
    \includegraphics[width=1\linewidth]{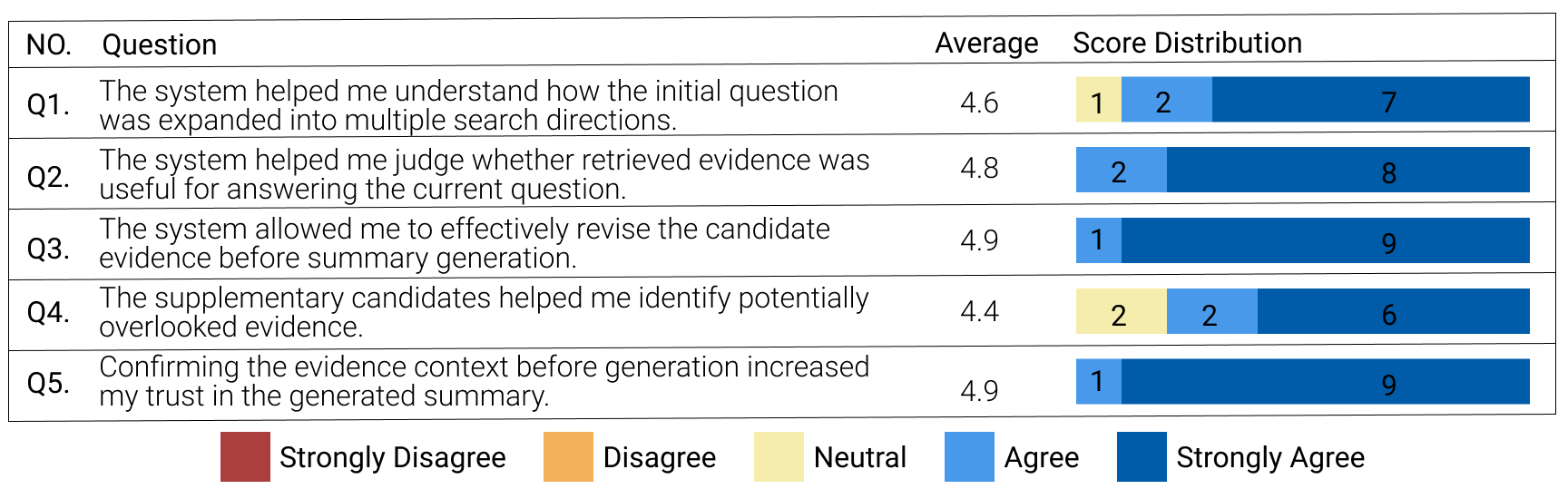}
    \caption{Subjective ratings on the effectiveness of \textit{FootprintRAG} for evidence context refinement.}
    \label{fig:user-study-rating}
    \vspace{-1.2em}
\end{figure}

\textbf{Subjective ratings.}
Figure~\ref{fig:user-study-rating} summarizes the questionnaire results. Overall, participants rated \textit{FootprintRAG} positively across all five questions, with all responses at or above neutral. Candidate evidence revision (Q3, \(M=4.9, SD=0.32\)) and evidence context confirmation (Q5, \(M=4.9, SD=0.32\)) received the highest ratings, with nine participants giving the highest score for each. Evidence usefulness judgment was also rated highly (Q2, \(M=4.8, SD=0.42\)), suggesting that participants found the system useful for judging whether retrieved evidence supported the task.

Query expansion visibility received positive ratings as well (Q1, \(M=4.6, SD=0.70\)), although one participant gave a neutral score. Supplementary candidate usefulness showed the largest variance (Q4, \(M=4.4, SD=0.84\)), indicating that supplementary candidates were helpful in some cases but varied more with the retrieval context. This result is consistent with our design decision to present supplementary candidates as optional evidence for inspection rather than automatically selected summary context.

\textbf{Interaction behavior.}
We analyzed participants' interaction logs to examine how they used \textit{FootprintRAG} during the exploration task. Evidence-level operations were the most frequent category (\(M=25.1, SD=10.90\) per participant), followed by retrieval-strategy adjustment (\(M=8.7, SD=4.42\)) and summary revision (\(M=3.7, SD=1.42\)). This pattern indicates that participants mainly worked with the evidence context itself, while also adjusting retrieval directions during exploration.

At the strategy level, participants used regeneration, replanning, and manual query input to redirect exploration paths. At the evidence level, they frequently removed reranked evidence units (\(M=10.6, SD=5.85\)) and added ERS-ranked supplementary candidates (\(M=12.9, SD=8.20\)), while local candidate additions were less frequent (\(M=1.6, SD=2.01\)). Summary-level operations were comparatively limited, especially whole-summary rewriting (\(M=0.3, SD=0.48\)). These results suggest that participants primarily refined retrieval directions and evidence contexts before making targeted summary revisions, rather than repeatedly rewriting the final output.


\textbf{Interview feedback.}

We organized participants' feedback into three themes. First, participants found the retrieval process more legible: \(P_2\), \(P_4\), \(P_7\), and \(P_9\) noted that the matrix showed how the system explored the corpus from different angles, and \(P_4\) and \(P_7\) said it helped them judge topic coverage.

Second, participants valued candidate-level intervention before generation. \(P_1\), \(P_3\), \(P_5\), \(P_6\), and \(P_{10}\) reported that retrieved candidates often mixed useful evidence with background descriptions, repeated tool introductions, or weakly related passages; candidate revision helped remove such evidence. \(P_3\) and \(P_6\) also noted that semantic relevance did not always imply evidence usefulness.

Third, participants considered context confirmation important for trust. \(P_1\), \(P_2\), \(P_5\), \(P_8\), and \(P_{10}\) reported that seeing selected evidence before generation made the final summary easier to inspect and justify, and \(P_5\) and \(P_8\) noted that the effort was acceptable when reliability and traceability mattered.

Overall, the questionnaire, logs, and interviews consistently show that participants used \textit{FootprintRAG} to inspect retrieval directions, revise weak or redundant evidence, and confirm evidence contexts before final synthesis. Supplementary candidates were useful in some cases, but their value depended on the retrieval context.

\section{Discussion}
\subsection{Multi-Perspective Workflow Comparison}
\label{sec:workflow-comparison}

To position \textit{FootprintRAG} within the broader landscape of RAG-based systems, we compare it with representative RAG-based systems and visual analytics tools from a workflow perspective.
As shown in Table~\ref{tab:workflow-comparison}, existing RAG systems mainly emphasize automatic context construction, multimodal retrieval, or developer-facing workflow inspection. \textit{FootprintRAG} complements them by treating generation context as user-refined evidence: it exposes candidate evidence before synthesis, supports revision across retrieval rounds, and grounds summaries in the confirmed context.

\begin{table*}[t]
\centering
\caption{Workflow-level comparison of representative RAG systems and visual analytics tools.}
\label{tab:workflow-comparison}
\footnotesize
\setlength{\tabcolsep}{4pt}
\renewcommand{\arraystretch}{1.18}
\begin{tabularx}{\textwidth}{
>{\raggedright\arraybackslash}p{2.15cm}
>{\raggedright\arraybackslash}p{2.85cm}
>{\raggedright\arraybackslash}p{2.75cm}
>{\raggedright\arraybackslash}p{3.05cm}
>{\raggedright\arraybackslash}X
}
\toprule
\textbf{System} 
& \textbf{Context Formation} 
& \textbf{Evidence Representation} 
& \textbf{Retrieval Strategy} 
& \textbf{Workflow Support} \\
\midrule

\textit{NotebookRAG}~\cite{shan2026notebookrag}
& Context for notebook generation
& Executable code cells
& Agent-guided retrieval for code generation
& No visual evidence refinement \\

\textit{ResearchAgent}~\cite{baek2025researchagent}
& Agent-constructed research context
& Literature passages
& Agent-driven iterative retrieval
& No visual workflow representation \\

\textit{VisRAG}~\cite{yu2025visrag}
& Top-\(k\) document-image context
& Document images and layout
& Single-stage visual retrieval
& No user-guided refinement workflow \\

\textit{RAGExplorer}~\cite{tian2026ragexplorervisualanalyticscomparative}
& Context for pipeline diagnosis
& Text chunks
& Retrieval configuration comparison
& Visual diagnosis of RAG configurations \\

\textit{RAGTrace}~\cite{cheng2025ragtrace}
& Post-hoc execution context
& Retrieved textual evidence
& Retrieval--generation tracing
& Visual analysis of failure paths \\

\textbf{FootprintRAG}
& \textbf{User-refined evidence context}
& \textbf{Textual and figure-related evidence units}
& \textbf{Multi-round retrieval with user revision}
& \textbf{Visual support for candidate and context refinement} \\

\bottomrule
\end{tabularx}
\end{table*}

\subsection{Latency and Cost}

\textit{FootprintRAG} introduces more latency than single-pass RAG because it performs query expansion, multi-variant retrieval, assessment, supplementary candidate selection, and summary generation. This cost was acceptable in our studies because the tasks were analytical and bounded by configurable parameters such as retrieval rounds, query variants, and retained candidates. For larger deployments, cost can be reduced through cached embeddings, batched or lightweight assessment, adaptive stopping, and reserving stronger models for final synthesis.

\subsection{Limitations and Future Work}

Although our evaluation shows how \textit{FootprintRAG} supports evidence context refinement in several scenarios, it does not establish general effectiveness across domains or corpus sizes. The current studies use task-bounded corpora of 10--48 papers; larger corpora may require row collapsing, strategy clustering, support-level filtering, cached embeddings, and batched or lightweight assessment.

\textit{FootprintRAG} represents figure evidence through captions, surrounding text, and LLM-generated descriptions rather than cross-modal embeddings. This enables a unified refinement workflow, but may miss fine-grained visual information such as numerical trends, data distributions, and spatial relationships. Future work could integrate cross-modal embeddings, chart-specific parsing, and visual verification.

The system also relies on LLM agents for query rewriting, candidate assessment, figure description, and summary generation, which may be sensitive to prompts and model choices. Future work should evaluate prompt/model sensitivity, surface uncertainty cues, and incorporate citation networks, publication time, and venue signals into supplementary evidence recommendation.

\section{Conclusion}

We presented \textit{FootprintRAG}, an LLM-agent-powered visual analytics system for evidence context refinement in RAG-based scientific literature exploration. The system focuses on the RAG stage where retrieved evidence is assessed, supplemented, revised, and confirmed before synthesis. Based on a formative study, we developed an evidence context refinement workflow that constructs text and figure evidence units, expands queries into multiple retrieval directions, assesses reranked units, and surfaces additional evidence through local candidate pools and ERS-ranked corpus-level recommendations. \textit{FootprintRAG} visualizes this workflow through four coordinated views for interactive corpus configuration, retrieval-pathway comparison, evidence-state revision, and evidence-grounded summary generation. Through two case studies, a user study, and a workflow-level comparison, we examined how \textit{FootprintRAG} helps users explore retrieval directions, revise candidate evidence, recover supplementary evidence, and trace generated summaries back to supporting evidence units. Overall, this work shows how visual analytics can make RAG-derived evidence contexts explicit and revisable, shifting user intervention from post-generation correction to pre-generation evidence context refinement.

\bibliographystyle{abbrv-doi}

\bibliography{edit_paper}

\begin{thebibliography}{10}

\bibitem{asai2026synthesizing}
A.~Asai, J.~He, R.~Shao, W.~Shi, A.~Singh, J.~C. Chang, K.~Lo, L.~Soldaini,
  S.~Feldman, M.~D'Arcy, D.~Wadden, M.~Latzke, J.~Sparks, J.~D. Hwang,
  V.~Kishore, M.~Tian, P.~Ji, S.~Liu, H.~Tong, B.~Wu, Y.~Xiong, L.~Zettlemoyer,
  G.~Neubig, D.~S. Weld, D.~Downey, W.-t. Yih, P.~W. Koh, and H.~Hajishirzi.
\newblock Synthesizing scientific literature with retrieval-augmented language
  models.
\newblock {\em Nature}, 650(8103):857--863, 2026. doi: {{%
10\hspace{.1pt}\discretionary{.}{%
}{.}\hspace{.4pt}1038\discretionary{/}{%
}{/}s41586\discretionary{%
}{-}{-}025\discretionary{%
}{-}{-}10072\discretionary{%
}{-}{-}4}}


\bibitem{baek2025researchagent}
J.~Baek, S.~K. Jauhar, S.~Cucerzan, and S.~J. Hwang.
\newblock {ResearchAgent}: Iterative research idea generation over scientific
  literature with large language models.
\newblock In {\em Proceedings of the 2025 Conference of the Nations of the
  Americas Chapter of the Association for Computational Linguistics: Human
  Language Technologies (Volume 1: Long Papers)}, pp. 6709--6738, 2025. doi:
  {{%
10\hspace{.1pt}\discretionary{.}{%
}{.}\hspace{.4pt}18653\discretionary{/}{%
}{/}v1\discretionary{/}{%
}{/}2025\hspace{.1pt}\discretionary{.}{%
}{.}\hspace{.4pt}naacl\discretionary{%
}{-}{-}long\hspace{.1pt}\discretionary{.}{%
}{.}\hspace{.4pt}342}}


\bibitem{beck2024puresuggest}
F.~Beck.
\newblock {PUREsuggest}: Citation-based literature search and visual
  exploration with keyword-controlled rankings.
\newblock {\em IEEE Transactions on Visualization and Computer Graphics},
  31(1):316--326, 2025. doi: {{%
10\hspace{.1pt}\discretionary{.}{%
}{.}\hspace{.4pt}1109\discretionary{/}{%
}{/}tvcg\hspace{.1pt}\discretionary{.}{%
}{.}\hspace{.4pt}2024\hspace{.1pt}\discretionary{.}{%
}{.}\hspace{.4pt}3456199}}


\bibitem{behera2023visual}
P.~K. Behera, S.~J. Jain, and A.~Kumar.
\newblock Visual exploration of literature using {Connected Papers}: A
  practical approach.
\newblock {\em Issues in Science and Technology Librarianship}, (104), 2023.
  doi: {{%
10\hspace{.1pt}\discretionary{.}{%
}{.}\hspace{.4pt}29173\discretionary{/}{%
}{/}istl2760}}


\bibitem{brown2025systematic}
A.~Brown, M.~Roman, and B.~Devereux.
\newblock A systematic literature review of retrieval-augmented generation:
  Techniques, metrics, and challenges.
\newblock {\em Big Data and Cognitive Computing}, 9(12):320, 2025. doi: {{%
10\hspace{.1pt}\discretionary{.}{%
}{.}\hspace{.4pt}3390\discretionary{/}{%
}{/}bdcc9120320}}


\bibitem{chang2023citesee}
J.~C. Chang, A.~X. Zhang, J.~Bragg, A.~Head, K.~Lo, D.~Downey, and D.~S. Weld.
\newblock {CiteSee}: Augmenting citations in scientific papers with persistent
  and personalized historical context.
\newblock In {\em Proceedings of the 2023 CHI Conference on Human Factors in
  Computing Systems}, pp. 1--15, 2023. doi: {{%
10\hspace{.1pt}\discretionary{.}{%
}{.}\hspace{.4pt}1145\discretionary{/}{%
}{/}3544548\hspace{.1pt}\discretionary{.}{%
}{.}\hspace{.4pt}3580847}}


\bibitem{chen2021vis30k}
J.~Chen, M.~Ling, R.~Li, P.~Isenberg, T.~Isenberg, M.~Sedlmair,
  T.~M{\"{o}}ller, R.~S. Laramee, H.-W. Shen, K.~W{\"{u}}nsche, and Q.~Wang.
\newblock {VIS30K}: A collection of figures and tables from {IEEE}
  visualization conference publications.
\newblock {\em IEEE Transactions on Visualization and Computer Graphics},
  27(9):3826--3833, 2021. doi: {{%
10\hspace{.1pt}\discretionary{.}{%
}{.}\hspace{.4pt}1109\discretionary{/}{%
}{/}tvcg\hspace{.1pt}\discretionary{.}{%
}{.}\hspace{.4pt}2021\hspace{.1pt}\discretionary{.}{%
}{.}\hspace{.4pt}3054916}}


\bibitem{chen2026keditvis}
Z.~Chen, H.~Zhan, Y.~Huang, X.~Wu, D.~Deng, D.~Weng, and Y.~Wu.
\newblock {KEditVis}: A visual analytics system for knowledge editing of large
  language models.
\newblock {\em IEEE Transactions on Visualization and Computer Graphics},
  32(6):4818--4828, 2026. doi: {{%
10\hspace{.1pt}\discretionary{.}{%
}{.}\hspace{.4pt}1109\discretionary{/}{%
}{/}tvcg\hspace{.1pt}\discretionary{.}{%
}{.}\hspace{.4pt}2026\hspace{.1pt}\discretionary{.}{%
}{.}\hspace{.4pt}3694436}}


\bibitem{cheng2025survey}
M.~Cheng, Y.~Luo, J.~Ouyang, Q.~Liu, H.~Liu, L.~Li, S.~Yu, B.~Zhang, J.~Cao,
  J.~Ma, D.~Wang, and E.~Chen.
\newblock A survey on knowledge-oriented retrieval-augmented generation.
\newblock {\em ACM Transactions on Information Systems}, 2026.
\newblock Advance online publication. doi: {{%
10\hspace{.1pt}\discretionary{.}{%
}{.}\hspace{.4pt}1145\discretionary{/}{%
}{/}3833415}}


\bibitem{cheng2025ragtrace}
S.~Cheng, J.~Li, H.~Wang, and Y.~Ma.
\newblock {RAGTrace}: Understanding and refining retrieval-generation dynamics
  in retrieval-augmented generation.
\newblock In {\em Proceedings of the 38th Annual ACM Symposium on User
  Interface Software and Technology}, pp. 1--20, 2025. doi: {{%
10\hspace{.1pt}\discretionary{.}{%
}{.}\hspace{.4pt}1145\discretionary{/}{%
}{/}3746059\hspace{.1pt}\discretionary{.}{%
}{.}\hspace{.4pt}3747741}}


\bibitem{ding2026scirag}
H.~Ding, Y.~Zhao, T.~Hu, M.~Patwardhan, and A.~Cohan.
\newblock {SciRAG}: Adaptive, citation-aware, and outline-guided retrieval and
  synthesis for scientific literature.
\newblock In {\em Proceedings of the 19th Conference of the European Chapter of
  the Association for Computational Linguistics (Volume 1: Long Papers)}, pp.
  6440--6460, 2026. doi: {{%
10\hspace{.1pt}\discretionary{.}{%
}{.}\hspace{.4pt}18653\discretionary{/}{%
}{/}v1\discretionary{/}{%
}{/}2026\hspace{.1pt}\discretionary{.}{%
}{.}\hspace{.4pt}eacl\discretionary{%
}{-}{-}long\hspace{.1pt}\discretionary{.}{%
}{.}\hspace{.4pt}303}}


\bibitem{yu2023user}
Y.~Dong, I.~Oppermann, J.~Liang, X.~Yuan, and Q.~V. Nguyen.
\newblock User-centered visual explorer of in-process comparison in
  spatiotemporal space.
\newblock {\em Journal of Visualization}, 26(2):403--421, 2023. doi: {{%
10\hspace{.1pt}\discretionary{.}{%
}{.}\hspace{.4pt}1007\discretionary{/}{%
}{/}s12650\discretionary{%
}{-}{-}022\discretionary{%
}{-}{-}00882\discretionary{%
}{-}{-}3}}


\bibitem{fok2023scim}
R.~Fok, H.~Kambhamettu, L.~Soldaini, J.~Bragg, K.~Lo, M.~Hearst, A.~Head, and
  D.~S. Weld.
\newblock {Scim}: Intelligent skimming support for scientific papers.
\newblock In {\em Proceedings of the 28th International Conference on
  Intelligent User Interfaces}, pp. 476--490, 2023. doi: {{%
10\hspace{.1pt}\discretionary{.}{%
}{.}\hspace{.4pt}1145\discretionary{/}{%
}{/}3581641\hspace{.1pt}\discretionary{.}{%
}{.}\hspace{.4pt}3584034}}


\bibitem{gao2023retrieval}
Y.~Gao, Y.~Xiong, X.~Gao, K.~Jia, J.~Pan, Y.~Bi, Y.~Dai, J.~Sun, M.~Wang, and
  H.~Wang.
\newblock Retrieval-augmented generation for large language models: A survey.
\newblock arXiv preprint arXiv:2312.10997, 2023. doi: {{%
10\hspace{.1pt}\discretionary{.}{%
}{.}\hspace{.4pt}48550\discretionary{/}{%
}{/}arXiv\hspace{.1pt}\discretionary{.}{%
}{.}\hspace{.4pt}2312\hspace{.1pt}\discretionary{.}{%
}{.}\hspace{.4pt}10997}}


\bibitem{ghafarollahi2025sciagents}
A.~Ghafarollahi and M.~J. Buehler.
\newblock {SciAgents}: Automating scientific discovery through bioinspired
  multi-agent intelligent graph reasoning.
\newblock {\em Advanced Materials}, 37(22):2413523, 2025. doi: {{%
10\hspace{.1pt}\discretionary{.}{%
}{.}\hspace{.4pt}1002\discretionary{/}{%
}{/}adma\hspace{.1pt}\discretionary{.}{%
}{.}\hspace{.4pt}202413523}}


\bibitem{gokdemir2025hiperrag}
O.~Gokdemir, C.~Siebenschuh, A.~Brace, A.~Wells, B.~Hsu, K.~Hippe, P.~Setty,
  A.~Ajith, J.~G. Pauloski, V.~Sastry, S.~Foreman, H.~Zheng, H.~Ma, B.~Kale,
  N.~Chia, T.~Gibbs, M.~Papka, T.~Brettin, F.~Alexander, A.~Anandkumar,
  I.~Foster, R.~Stevens, V.~Vishwanath, and A.~Ramanathan.
\newblock {HiPerRAG}: High-performance retrieval augmented generation for
  scientific insights.
\newblock In {\em Proceedings of the Platform for Advanced Scientific Computing
  Conference}, pp. 1--13, 2025. doi: {{%
10\hspace{.1pt}\discretionary{.}{%
}{.}\hspace{.4pt}1145\discretionary{/}{%
}{/}3732775\hspace{.1pt}\discretionary{.}{%
}{.}\hspace{.4pt}3733586}}


\bibitem{han2025fine}
J.~Han, Z.~Mao, Y.~Liu, Y.~Che, Z.~Fu, and Q.~Wang.
\newblock Fine-grained knowledge enhancement for retrieval-augmented
  generation.
\newblock In {\em Findings of the Association for Computational Linguistics:
  ACL 2025}, pp. 10031--10044, 2025. doi: {{%
10\hspace{.1pt}\discretionary{.}{%
}{.}\hspace{.4pt}18653\discretionary{/}{%
}{/}v1\discretionary{/}{%
}{/}2025\hspace{.1pt}\discretionary{.}{%
}{.}\hspace{.4pt}findings\discretionary{%
}{-}{-}acl\hspace{.1pt}\discretionary{.}{%
}{.}\hspace{.4pt}522}}


\bibitem{hwang2025retrieval}
J.~Hwang, J.~Park, H.~Park, D.~Kim, S.~Park, and J.~Ok.
\newblock Retrieval-augmented generation with estimation of source reliability.
\newblock In {\em Proceedings of the 2025 Conference on Empirical Methods in
  Natural Language Processing}, pp. 34267--34291, 2025. doi: {{%
10\hspace{.1pt}\discretionary{.}{%
}{.}\hspace{.4pt}18653\discretionary{/}{%
}{/}v1\discretionary{/}{%
}{/}2025\hspace{.1pt}\discretionary{.}{%
}{.}\hspace{.4pt}emnlp\discretionary{%
}{-}{-}main\hspace{.1pt}\discretionary{.}{%
}{.}\hspace{.4pt}1738}}


\bibitem{jiang2023graphologue}
P.~Jiang, J.~Rayan, S.~P. Dow, and H.~Xia.
\newblock {Graphologue}: Exploring large language model responses with
  interactive diagrams.
\newblock In {\em Proceedings of the 36th Annual ACM Symposium on User
  Interface Software and Technology}, pp. 1--20, 2023. doi: {{%
10\hspace{.1pt}\discretionary{.}{%
}{.}\hspace{.4pt}1145\discretionary{/}{%
}{/}3586183\hspace{.1pt}\discretionary{.}{%
}{.}\hspace{.4pt}3606737}}


\bibitem{kalra2024hypa}
R.~Kalra, Z.~Wu, A.~Gulley, A.~Hilliard, X.~Guan, A.~Koshiyama, and P.~C.
  Treleaven.
\newblock {HyPA-RAG}: A hybrid parameter adaptive retrieval-augmented
  generation system for {AI} legal and policy applications.
\newblock In {\em Proceedings of the 1st Workshop on Customizable NLP: Progress
  and Challenges in Customizing NLP for a Domain, Application, Group, or
  Individual (CustomNLP4U)}, pp. 237--256, 2024. doi: {{%
10\hspace{.1pt}\discretionary{.}{%
}{.}\hspace{.4pt}18653\discretionary{/}{%
}{/}v1\discretionary{/}{%
}{/}2024\hspace{.1pt}\discretionary{.}{%
}{.}\hspace{.4pt}customnlp4u\discretionary{%
}{-}{-}1\hspace{.1pt}\discretionary{.}{%
}{.}\hspace{.4pt}18}}


\bibitem{kang2022threddy}
H.~Kang, J.~C. Chang, Y.~Kim, and A.~Kittur.
\newblock {Threddy}: An interactive system for personalized thread-based
  exploration and organization of scientific literature.
\newblock In {\em Proceedings of the 35th Annual ACM Symposium on User
  Interface Software and Technology}, pp. 1--15, 2022. doi: {{%
10\hspace{.1pt}\discretionary{.}{%
}{.}\hspace{.4pt}1145\discretionary{/}{%
}{/}3526113\hspace{.1pt}\discretionary{.}{%
}{.}\hspace{.4pt}3545660}}


\bibitem{lee2025hybgrag}
M.-C. Lee, Q.~Zhu, C.~Mavromatis, Z.~Han, S.~Adeshina, V.~N. Ioannidis,
  H.~Rangwala, and C.~Faloutsos.
\newblock {HybGRAG}: Hybrid retrieval-augmented generation on textual and
  relational knowledge bases.
\newblock In {\em Proceedings of the 63rd Annual Meeting of the Association for
  Computational Linguistics (Volume 1: Long Papers)}, pp. 879--893, 2025. doi:
  {{%
10\hspace{.1pt}\discretionary{.}{%
}{.}\hspace{.4pt}18653\discretionary{/}{%
}{/}v1\discretionary{/}{%
}{/}2025\hspace{.1pt}\discretionary{.}{%
}{.}\hspace{.4pt}acl\discretionary{%
}{-}{-}long\hspace{.1pt}\discretionary{.}{%
}{.}\hspace{.4pt}43}}


\bibitem{lee2024paperweaver}
Y.~Lee, H.~B. Kang, M.~Latzke, J.~Kim, J.~Bragg, J.~C. Chang, and
  P.~Siangliulue.
\newblock {PaperWeaver}: Enriching topical paper alerts by contextualizing
  recommended papers with user-collected papers.
\newblock In {\em Proceedings of the CHI Conference on Human Factors in
  Computing Systems}, pp. 1--19, 2024. doi: {{%
10\hspace{.1pt}\discretionary{.}{%
}{.}\hspace{.4pt}1145\discretionary{/}{%
}{/}3613904\hspace{.1pt}\discretionary{.}{%
}{.}\hspace{.4pt}3642196}}


\bibitem{li2025reasongraph}
Z.~Li, E.~Shareghi, and N.~Collier.
\newblock {ReasonGraph}: Visualization of reasoning methods and extended
  inference paths.
\newblock In {\em Proceedings of the 63rd Annual Meeting of the Association for
  Computational Linguistics (Volume 3: System Demonstrations)}, pp. 140--147,
  2025. doi: {{%
10\hspace{.1pt}\discretionary{.}{%
}{.}\hspace{.4pt}18653\discretionary{/}{%
}{/}v1\discretionary{/}{%
}{/}2025\hspace{.1pt}\discretionary{.}{%
}{.}\hspace{.4pt}acl\discretionary{%
}{-}{-}demo\hspace{.1pt}\discretionary{.}{%
}{.}\hspace{.4pt}14}}


\bibitem{liu2025city}
G.~Liu, Y.~Jiang, X.~Yan, N.~Cao, and Y.~Shi.
\newblock {City of Wander}: Visualizing scientific literature for knowledge
  exploration using visual metaphors.
\newblock In {\em Proceedings of the Extended Abstracts of the CHI Conference
  on Human Factors in Computing Systems}, pp. 1--7, 2025. doi: {{%
10\hspace{.1pt}\discretionary{.}{%
}{.}\hspace{.4pt}1145\discretionary{/}{%
}{/}3706599\hspace{.1pt}\discretionary{.}{%
}{.}\hspace{.4pt}3720280}}


\bibitem{lu2024agentlens}
J.~Lu, B.~Pan, J.~Chen, Y.~Feng, J.~Hu, Y.~Peng, and W.~Chen.
\newblock {AgentLens}: Visual analysis for agent behaviors in {LLM}-based
  autonomous systems.
\newblock {\em IEEE Transactions on Visualization and Computer Graphics},
  31(8):4182--4197, 2025. doi: {{%
10\hspace{.1pt}\discretionary{.}{%
}{.}\hspace{.4pt}1109\discretionary{/}{%
}{/}tvcg\hspace{.1pt}\discretionary{.}{%
}{.}\hspace{.4pt}2024\hspace{.1pt}\discretionary{.}{%
}{.}\hspace{.4pt}3394053}}


\bibitem{lu2026agentic}
X.~Lu, G.~Li, Y.~Dong, R.~Peng, Z.~Wang, D.~Tian, and G.~Shan.
\newblock Agentic scientific visualization generation and caption semantic
  alignment.
\newblock {\em Information Visualization}, 25(3):213--231, 2026. doi: {{%
10\hspace{.1pt}\discretionary{.}{%
}{.}\hspace{.4pt}1177\discretionary{/}{%
}{/}14738716261434841}}


\bibitem{bran2023chemcrow}
A.~M.~Bran, S.~Cox, O.~Schilter, C.~Baldassari, A.~D. White, and P.~Schwaller.
\newblock Augmenting large language models with chemistry tools.
\newblock {\em Nature Machine Intelligence}, 6(5):525--535, 2024. doi: {{%
10\hspace{.1pt}\discretionary{.}{%
}{.}\hspace{.4pt}1038\discretionary{/}{%
}{/}s42256\discretionary{%
}{-}{-}024\discretionary{%
}{-}{-}00832\discretionary{%
}{-}{-}8}}


\bibitem{narechania2021vitality}
A.~Narechania, A.~Karduni, R.~Wesslen, and E.~Wall.
\newblock {VITALITY}: Promoting serendipitous discovery of academic literature
  with transformers \& visual analytics.
\newblock {\em IEEE Transactions on Visualization and Computer Graphics},
  28(1):486--496, 2022. doi: {{%
10\hspace{.1pt}\discretionary{.}{%
}{.}\hspace{.4pt}1109\discretionary{/}{%
}{/}tvcg\hspace{.1pt}\discretionary{.}{%
}{.}\hspace{.4pt}2021\hspace{.1pt}\discretionary{.}{%
}{.}\hspace{.4pt}3114820}}


\bibitem{niu2025mineru2}
J.~Niu, Z.~Liu, Z.~Gu, B.~Wang, L.~Ouyang, Z.~Zhao, T.~Chu, T.~He, F.~Wu,
  Q.~Zhang, Z.~Jin, G.~Liang, R.~Zhang, W.~Zhang, Y.~Qu, Z.~Ren, Y.~Sun,
  Z.~Tang, B.~Niu, Y.~Zheng, D.~Ma, Z.~Miao, H.~Dong, S.~Qian, J.~Zhang,
  F.~Wang, J.~Chen, X.~Zhao, L.~Wei, W.~Li, S.~Wang, R.~Xu, Y.~Cao, L.~Chen,
  Q.~Wu, H.~Gu, L.~Lu, D.~Lin, G.~Shen, X.~Zhou, L.~Zhang, Y.~Zang, X.~Dong,
  J.~Wang, B.~Zhang, L.~Bai, P.~Chu, W.~Li, J.~Wu, L.~Wu, Z.~Li, G.~Wang,
  Z.~Tu, C.~Xu, K.~Chen, B.~Zhou, D.~Lin, W.~Zhang, and C.~He.
\newblock {MinerU2.5}: A decoupled vision-language model for efficient
  high-resolution document parsing.
\newblock In {\em Proceedings of the 64th Annual Meeting of the Association for
  Computational Linguistics (Volume 6: Industry Track)}, pp. 13--42, 2026. doi:
  {{%
10\hspace{.1pt}\discretionary{.}{%
}{.}\hspace{.4pt}18653\discretionary{/}{%
}{/}v1\discretionary{/}{%
}{/}2026\hspace{.1pt}\discretionary{.}{%
}{.}\hspace{.4pt}acl\discretionary{%
}{-}{-}industry\hspace{.1pt}\discretionary{.}{%
}{.}\hspace{.4pt}3}}


\bibitem{oche2025systematic}
A.~J. Oche, A.~G. Folashade, T.~Ghosal, and A.~Biswas.
\newblock A systematic review of key retrieval-augmented generation ({RAG})
  systems: Progress, gaps, and future directions.
\newblock arXiv preprint arXiv:2507.18910, 2025. doi: {{%
10\hspace{.1pt}\discretionary{.}{%
}{.}\hspace{.4pt}48550\discretionary{/}{%
}{/}arXiv\hspace{.1pt}\discretionary{.}{%
}{.}\hspace{.4pt}2507\hspace{.1pt}\discretionary{.}{%
}{.}\hspace{.4pt}18910}}


\bibitem{palani2023relatedly}
S.~Palani, A.~Naik, D.~Downey, A.~X. Zhang, J.~Bragg, and J.~C. Chang.
\newblock {Relatedly}: Scaffolding literature reviews with existing related
  work sections.
\newblock In {\em Proceedings of the 2023 CHI Conference on Human Factors in
  Computing Systems}, pp. 1--20, 2023. doi: {{%
10\hspace{.1pt}\discretionary{.}{%
}{.}\hspace{.4pt}1145\discretionary{/}{%
}{/}3544548\hspace{.1pt}\discretionary{.}{%
}{.}\hspace{.4pt}3580841}}


\bibitem{pang2026interactive}
R.~Y. Pang, K.~J.~K. Feng, S.~Feng, C.~Li, W.~Shi, Y.~Tsvetkov, J.~Heer, and
  K.~Reinecke.
\newblock {Interactive Reasoning}: Visualizing and controlling chain-of-thought
  reasoning in large language models.
\newblock In {\em Proceedings of the 31st International Conference on
  Intelligent User Interfaces}, pp. 852--867, 2026. doi: {{%
10\hspace{.1pt}\discretionary{.}{%
}{.}\hspace{.4pt}1145\discretionary{/}{%
}{/}3742413\hspace{.1pt}\discretionary{.}{%
}{.}\hspace{.4pt}3789091}}


\bibitem{peng2025graph}
B.~Peng, Y.~Zhu, Y.~Liu, X.~Bo, H.~Shi, C.~Hong, Y.~Zhang, and S.~Tang.
\newblock Graph retrieval-augmented generation: A survey.
\newblock {\em ACM Transactions on Information Systems}, 44(2):1--52, 2026.
  doi: {{%
10\hspace{.1pt}\discretionary{.}{%
}{.}\hspace{.4pt}1145\discretionary{/}{%
}{/}3777378}}


\bibitem{peng2025textlens}
R.~Peng, Y.~Dong, G.~Li, D.~Tian, and G.~Shan.
\newblock {TextLens}: large language models-powered visual analytics enhancing
  text clustering.
\newblock {\em Journal of Visualization}, 28(3):625--643, 2025. doi: {{%
10\hspace{.1pt}\discretionary{.}{%
}{.}\hspace{.4pt}1007\discretionary{/}{%
}{/}s12650\discretionary{%
}{-}{-}025\discretionary{%
}{-}{-}01043\discretionary{%
}{-}{-}y}}


\bibitem{qiu2022docflow}
R.~Qiu, Y.~Tu, Y.-S. Wang, P.-Y. Yen, and H.-W. Shen.
\newblock {DocFlow}: A visual analytics system for question-based document
  retrieval and categorization.
\newblock {\em IEEE Transactions on Visualization and Computer Graphics},
  30(2):1533--1548, 2024. doi: {{%
10\hspace{.1pt}\discretionary{.}{%
}{.}\hspace{.4pt}1109\discretionary{/}{%
}{/}tvcg\hspace{.1pt}\discretionary{.}{%
}{.}\hspace{.4pt}2022\hspace{.1pt}\discretionary{.}{%
}{.}\hspace{.4pt}3219762}}


\bibitem{shan2026notebookrag}
Y.~Shan, Y.~He, Z.~Shao, K.~Xu, and S.~Chen.
\newblock {NotebookRAG}: Retrieving multiple notebooks to augment the
  generation of {EDA} notebooks for crowd-wisdom.
\newblock In {\em 2026 IEEE 19th Pacific Visualization Conference
  (PacificVis)}, pp. 346--356, 2026. doi: {{%
10\hspace{.1pt}\discretionary{.}{%
}{.}\hspace{.4pt}1109\discretionary{/}{%
}{/}pacificvis68791\hspace{.1pt}\discretionary{.}{%
}{.}\hspace{.4pt}2026\hspace{.1pt}\discretionary{.}{%
}{.}\hspace{.4pt}00043}}


\bibitem{sharma2025retrieval}
C.~Sharma.
\newblock Retrieval-augmented generation: A comprehensive survey of
  architectures, enhancements, and robustness frontiers.
\newblock arXiv preprint arXiv:2506.00054, 2025. doi: {{%
10\hspace{.1pt}\discretionary{.}{%
}{.}\hspace{.4pt}48550\discretionary{/}{%
}{/}arXiv\hspace{.1pt}\discretionary{.}{%
}{.}\hspace{.4pt}2506\hspace{.1pt}\discretionary{.}{%
}{.}\hspace{.4pt}00054}}


\bibitem{tian2026ragexplorervisualanalyticscomparative}
H.~Tian, Y.~Feng, Z.~Wen, H.~Li, M.~Zhu, and W.~Chen.
\newblock {RAGExplorer}: A visual analytics system for the comparative
  diagnosis of {RAG} systems.
\newblock {\em IEEE Transactions on Visualization and Computer Graphics},
  32(6):4807--4817, 2026. doi: {{%
10\hspace{.1pt}\discretionary{.}{%
}{.}\hspace{.4pt}1109\discretionary{/}{%
}{/}tvcg\hspace{.1pt}\discretionary{.}{%
}{.}\hspace{.4pt}2026\hspace{.1pt}\discretionary{.}{%
}{.}\hspace{.4pt}3694443}}


\bibitem{tian2023litvis}
M.~Tian, G.~Li, and X.~Yuan.
\newblock {LitVis}: a visual analytics approach for managing and exploring
  literature.
\newblock {\em Journal of Visualization}, 26(6):1445--1458, 2023. doi: {{%
10\hspace{.1pt}\discretionary{.}{%
}{.}\hspace{.4pt}1007\discretionary{/}{%
}{/}s12650\discretionary{%
}{-}{-}023\discretionary{%
}{-}{-}00941\discretionary{%
}{-}{-}3}}


\bibitem{tu2022phrasemap}
Y.~Tu, R.~Qiu, Y.-S. Wang, P.-Y. Yen, and H.-W. Shen.
\newblock {PhraseMap}: Attention-based keyphrases recommendation for
  information seeking.
\newblock {\em IEEE Transactions on Visualization and Computer Graphics},
  30(3):1787--1802, 2024. doi: {{%
10\hspace{.1pt}\discretionary{.}{%
}{.}\hspace{.4pt}1109\discretionary{/}{%
}{/}tvcg\hspace{.1pt}\discretionary{.}{%
}{.}\hspace{.4pt}2022\hspace{.1pt}\discretionary{.}{%
}{.}\hspace{.4pt}3225114}}


\bibitem{wang2025xgraphrag}
K.~Wang, B.~Pan, Y.~Feng, Y.~Wu, J.~Chen, M.~Zhu, and W.~Chen.
\newblock {XGraphRAG}: Interactive visual analysis for graph-based
  retrieval-augmented generation.
\newblock In {\em 2025 IEEE 18th Pacific Visualization Conference
  (PacificVis)}, pp. 1--11, 2025. doi: {{%
10\hspace{.1pt}\discretionary{.}{%
}{.}\hspace{.4pt}1109\discretionary{/}{%
}{/}pacificvis64226\hspace{.1pt}\discretionary{.}{%
}{.}\hspace{.4pt}2025\hspace{.1pt}\discretionary{.}{%
}{.}\hspace{.4pt}00005}}


\bibitem{wang2024ragviz}
T.~Wang, J.~He, and C.~Xiong.
\newblock {RAGViz}: Diagnose and visualize retrieval-augmented generation.
\newblock In {\em Proceedings of the 2024 Conference on Empirical Methods in
  Natural Language Processing: System Demonstrations}, pp. 320--327, 2024. doi:
  {{%
10\hspace{.1pt}\discretionary{.}{%
}{.}\hspace{.4pt}18653\discretionary{/}{%
}{/}v1\discretionary{/}{%
}{/}2024\hspace{.1pt}\discretionary{.}{%
}{.}\hspace{.4pt}emnlp\discretionary{%
}{-}{-}demo\hspace{.1pt}\discretionary{.}{%
}{.}\hspace{.4pt}33}}


\bibitem{wang2024surveyagent}
X.~Wang, J.~Chen, N.~Li, L.~Chen, X.~Yuan, W.~Shi, X.~Ge, R.~Xu, and Y.~Xiao.
\newblock {SurveyAgent}: A conversational system for personalized and efficient
  research survey.
\newblock arXiv preprint arXiv:2404.06364, 2024. doi: {{%
10\hspace{.1pt}\discretionary{.}{%
}{.}\hspace{.4pt}48550\discretionary{/}{%
}{/}arXiv\hspace{.1pt}\discretionary{.}{%
}{.}\hspace{.4pt}2404\hspace{.1pt}\discretionary{.}{%
}{.}\hspace{.4pt}06364}}


\bibitem{wang2024searching}
X.~Wang, Z.~Wang, X.~Gao, F.~Zhang, Y.~Wu, Z.~Xu, T.~Shi, Z.~Wang, S.~Li,
  Q.~Qian, R.~Yin, C.~Lv, X.~Zheng, and X.~Huang.
\newblock Searching for best practices in retrieval-augmented generation.
\newblock In {\em Proceedings of the 2024 Conference on Empirical Methods in
  Natural Language Processing}, pp. 17716--17736, 2024. doi: {{%
10\hspace{.1pt}\discretionary{.}{%
}{.}\hspace{.4pt}18653\discretionary{/}{%
}{/}v1\discretionary{/}{%
}{/}2024\hspace{.1pt}\discretionary{.}{%
}{.}\hspace{.4pt}emnlp\discretionary{%
}{-}{-}main\hspace{.1pt}\discretionary{.}{%
}{.}\hspace{.4pt}981}}


\bibitem{wu2023vizoptics}
C.~Wu, Y.~Chen, Y.~Dong, F.~Zhou, Y.~Zhao, and C.~J. Liang.
\newblock {VizOPTICS}: Getting insights into {OPTICS} via interactive visual
  analysis.
\newblock {\em Computers and Electrical Engineering}, 107:108624, 2023. doi:
  {{%
10\hspace{.1pt}\discretionary{.}{%
}{.}\hspace{.4pt}1016\discretionary{/}{%
}{/}j\hspace{.1pt}\discretionary{.}{%
}{.}\hspace{.4pt}compeleceng\hspace{.1pt}\discretionary{.}{%
}{.}\hspace{.4pt}2023\hspace{.1pt}\discretionary{.}{%
}{.}\hspace{.4pt}108624}}


\bibitem{yan2024corrective}
S.-Q. Yan, J.-C. Gu, Y.~Zhu, and Z.-H. Ling.
\newblock Corrective retrieval augmented generation.
\newblock arXiv preprint arXiv:2401.15884, 2024. doi: {{%
10\hspace{.1pt}\discretionary{.}{%
}{.}\hspace{.4pt}48550\discretionary{/}{%
}{/}arXiv\hspace{.1pt}\discretionary{.}{%
}{.}\hspace{.4pt}2401\hspace{.1pt}\discretionary{.}{%
}{.}\hspace{.4pt}15884}}


\bibitem{yang2025litforager}
A.~Yang, E.~H. Faa, W.~Liu, S.~Guo, D.~H. Chau, and Y.~Yang.
\newblock {LitForager}: Exploring multimodal literature foraging strategies in
  immersive sensemaking.
\newblock {\em IEEE Transactions on Visualization and Computer Graphics},
  31(11):9614--9624, 2025. doi: {{%
10\hspace{.1pt}\discretionary{.}{%
}{.}\hspace{.4pt}1109\discretionary{/}{%
}{/}tvcg\hspace{.1pt}\discretionary{.}{%
}{.}\hspace{.4pt}2025\hspace{.1pt}\discretionary{.}{%
}{.}\hspace{.4pt}3616732}}


\bibitem{yu2025visrag}
S.~Yu, C.~Tang, B.~Xu, J.~Cui, J.~Ran, Y.~Yan, Z.~Liu, S.~Wang, X.~Han, Z.~Liu,
  and M.~Sun.
\newblock {VisRAG}: Vision-based retrieval-augmented generation on
  multi-modality documents.
\newblock In {\em The Thirteenth International Conference on Learning
  Representations}, 2025.

\bibitem{zhang2025auragenome}
C.~Zhang, Y.~Dong, Y.~Wang, Y.~Han, G.~Shan, and B.~Tang.
\newblock {AuraGenome}: An {LLM}-powered framework for on-the-fly reusable and
  scalable circular genome visualizations.
\newblock {\em IEEE Computer Graphics and Applications}, 45(5):78--92, 2025.
  doi: {{%
10\hspace{.1pt}\discretionary{.}{%
}{.}\hspace{.4pt}1109\discretionary{/}{%
}{/}mcg\hspace{.1pt}\discretionary{.}{%
}{.}\hspace{.4pt}2025\hspace{.1pt}\discretionary{.}{%
}{.}\hspace{.4pt}3581560}}


\bibitem{zhang2023concepteva}
X.~Zhang, J.~Li, P.-W. Chi, S.~Chandrasegaran, and K.-L. Ma.
\newblock {ConceptEVA}: Concept-based interactive exploration and customization
  of document summaries.
\newblock In {\em Proceedings of the 2023 CHI Conference on Human Factors in
  Computing Systems}, pp. 1--16, 2023. doi: {{%
10\hspace{.1pt}\discretionary{.}{%
}{.}\hspace{.4pt}1145\discretionary{/}{%
}{/}3544548\hspace{.1pt}\discretionary{.}{%
}{.}\hspace{.4pt}3581260}}


\bibitem{zhao2026toward}
G.~Zhao, Z.~Wang, Y.~Dong, G.~Li, and G.~Shan.
\newblock Toward reliable scientific visualization pipeline construction with
  structure-aware retrieval-augmented {LLMs}.
\newblock {\em Information Visualization}, 25(3):373--390, 2026. doi: {{%
10\hspace{.1pt}\discretionary{.}{%
}{.}\hspace{.4pt}1177\discretionary{/}{%
}{/}14738716261434848}}


\end{thebibliography}

\section*{Appendix -- LLM Prompt Gallery}

\begin{tcolorbox}[
  enhanced, breakable,
  title={Query Agent Prompt},
  colback=black!2, colframe=black!25,
  boxrule=0.4pt, arc=1pt,
  left=8pt, right=8pt, top=6pt, bottom=6pt
]
You are the Chief Scientist in a scientific discovery team.\\[2pt]
\textbf{Mission Overview:}
Your team's goal is to investigate literature deeply and broadly, starting from the user's question, and discover new, valuable, and verifiable scientific knowledge.

You are building a research evidence matrix from retrieved and reviewed papers.  You lead the direction of retrieval.Evaluators execute your retrieval plans, inspect text and figures, summarize findings, and assess whether each result should be expanded.

\textbf{Your Core Responsibilities}
           
\begin{enumerate}[leftmargin=1.2em,itemsep=2pt,label=\arabic*)]
    \item  Initial planning: for a new user question, propose \texttt{plan\_per\_round}  search angles.
    \item  Dynamic decisions: adapt using prior rounds' strategies and plan summaries (answer/suggestion per plan). Do not rely on raw evidence text in the prompt.
\end{enumerate}
\textbf{Decision Rules}

When continuing retrieval, output a JSON list of strategy objects.Each strategy object must have:
\begin{enumerate}[leftmargin=1.2em,itemsep=2pt,label=\arabic*)]
    \item  action: fixed as call\_tool
    \item  tool\_name: usually strategy\_semantic\_search, strategy\_metadata\_search or strategy\_exact\_search
    \item args: tool arguments
    \item reason: concise rationale
\end{enumerate}   

Search Methods:
\begin{enumerate}[leftmargin=1.2em,itemsep=2pt,label=\arabic*)]
    \item  strategy\_semantic\_search: Used for semantic similarity retrieval. Parameter: query\_intent (a natural language phrase or sentence describing what you are looking for).
    \item  strategy\_metadata /search: Used for retrieving a specific paper to explore its context. Parameter: paper\_id (the ID of the paper).
    \item  strategy\_exact\_search: Used for exact text matching in the database. Parameter: query\_intent. IMPORTANT: For exact search, you MUST use ONLY one or two specific proper nouns (e.g. "PM2.5" or "CNN-LSTM") rather than long phrases or sentences to prevent getting zero results. Prefer terms implied by prior plansummary.answer
\end{enumerate}

\textbf{Prior rounds}
Below is the previous rounds context of the scientific discovery. User question is \{self.graph.root\_goal\}. Every strategy must directly serve this question and avoid irrelevant drift.Completed planner-evaluation cycles so far: \{self.round\_count\}. 

You have made up this querys:\{self.history\_querys\}.
Now you are expected to plan a new query based on the query \{self.this\_query\}.It found these results:
Evidence unit from paper\{self.this\_results.\_paper\}.The content is \{self.this\_results.\_content\}.The assessment agent suggestion is \{self.this\_results.\_evaluation.\_suggestion\}

\textbf{Output format} 

    [

      \{

        "action": "call\_tool",

        "ParentNode": "0",

        "tool\_name": "strategy\_semantic\_search",

        "args": \{ "query\_intent": "PM2.5 chemical composition and source analysis" \},

        "reason": "Prior work already covers PM2.5 concentration trends, but chemical composition and sources are under-explored; we should retrieve studies on PM2.5 composition apportionment and source analysis for a fuller picture."

      \},

      \{

        "action": "call\_tool",

        "ParentNode": "chunk\_003329",

        "tool\_name": "strategy\_metadata\_search",

        "args": \{ "paper\_id": "3" \},

        "reason": "This hit discusses air-pollutant monitoring methods and was rated highly by the evaluator; opening the full paper should clarify methods and findings for follow-up validation."

      \},

      \{

        "action": "call\_tool",

        "ParentNode": "0",

        "tool\_name": "strategy\_exact\_search",

        "args": \{ "query\_intent": "VOC" \},

        "reason": "We need passages that explicitly mention VOC; exact match avoids overly broad semantic noise."

      \},

      \{

        "action": "call\_tool",

        "ParentNode": "chunk\_002431",

        "tool\_name": "strategy\_semantic\_search",

        "args": \{ "query\_intent": "Atmospheric dispersion model improvements and applications" \},

        "reason": "Results mention pollutant dispersion but not model advances; retrieving dispersion-model improvements should strengthen prediction-oriented evidence."

      \}

    ]

    Or when evidence is already sufficient:

    [ \{ "action": "finish", "reason": "Evidence is sufficient to answer the user question, so retrieval can stop." \} ]

\end{tcolorbox}

\begin{tcolorbox}[
  enhanced, breakable,
  title={Assessment Agent Prompt},
  colback=black!2, colframe=black!25,
  boxrule=0.4pt, arc=1pt,
  left=8pt, right=8pt, top=6pt, bottom=6pt
]
\textbf{Task Description}

You are an Evaluator in a scientific literature investigation team.
Your task is to evaluate each retrieved evidence item independently and provide structured feedback for planning.

Evaluation Logic

Read title, summary, and insight carefully.
If an image is available, include visual judgment.

\textbf{Evaluation}

GROW: high-value evidence, strongly relevant, with clear follow-up clues.Will be used to determine next research step.
KEEP: medium-value evidence, relevant but limited expansion value.Will be used to write research summary.
PRUNE: low-value evidence, irrelevant/noisy/redundant.

extracted\_insight: Provide a concrete scientific observation, not generic praise.

suggested\_keywords: Extract potentially useful technical terms for follow-up search.

\textbf{Output Format}

Return a JSON list only no Markdown.

[
\{

"target\_evidence\_id": "fig\_001",

"branch\_action": "GROW",

"extracted\_insight": "The figure shows a positive correlation between winter PM2.5 peaks and respiratory emergency visits.",

"scores":\{"relevance": 9, "credibility": 8\},

"reason": "High-value trend evidence with follow-up paper-level traceability.",

"suggested\_keywords": ["time-series analysis", "respiratory emergency visits"]

\}
]

\end{tcolorbox}

\begin{tcolorbox}[
  enhanced, breakable,
  title={Context Agent Prompt},
  colback=black!2, colframe=black!25,
  boxrule=0.4pt, arc=1pt,
  left=8pt, right=8pt, top=6pt, bottom=6pt
]
\textbf{Task Description}

You are a scientific report architect.

You will be given a list of query strategies and their summaries each with a plan\_id.
Group these strategies logically into a 2-level outline for a scientific review report.

When assigning plan\_ids, follow these rules:

\begin{enumerate}
    \item Each plan\_id should be assigned to the most appropriate subsection based on its main scientific contribution.
    \item Avoid assigning the same plan\_id to multiple subsections unless the strategy clearly supports multiple distinct scientific topics.
    \item Do not create empty subsections.
    \item Do not create overly fragmented subsections that contain only minor wording differences.
    \item Prefer a balanced structure in which each level-1 theme contains one or more meaningful level-2 subsections.
    \item If several strategies are redundant, group them together under a single subsection rather than creating repeated sections.
    \item If a strategy is weak, noisy, or only marginally relevant, place it under the closest relevant subsection rather than inventing an unrelated theme.
\end{enumerate}

\textbf{Output Format}

[
  \{
  
    "level1\_title": "Broad Theme 1",
    
    "subsections": 
    [
      \{
      
        "level2\_title": "Specific Topic 1",
        
        "assigned\_plan\_ids": ["plan\_id\_1", "plan\_id\_2"]
        
      \}
    ]
    
  \}
]

\end{tcolorbox}

\begin{tcolorbox}[
  enhanced, breakable,
  title={Context Agent},
  colback=black!2, colframe=black!25,
  boxrule=0.4pt, arc=1pt,
  left=8pt, right=8pt, top=6pt, bottom=6pt
]
\textbf{Task Description}

You are a scientific literature review writer specialized in evidence-based synthesis.

You will be given a subsection title and a set of retrieved evidence items. Each evidence item contains a unique CHUNK\_ID, along with its corresponding textual content, summary, insight, or other relevant metadata. Your task is to write a concise, coherent, and professional English literature review paragraph that synthesizes the provided evidence under the given subsection title.
The paragraph must contain 3 to 6 sentences. Each sentence should contribute meaningful synthesis rather than repetition.

You MUST cite the retrieved evidence using EXACTLY the format [CHUNK\_ID], where CHUNK\_ID is replaced by the original evidence identifier.

Use citations wherever a claim is supported by a specific evidence item. Multiple citations may be used in the same sentence when the sentence synthesizes findings from multiple evidence items.

\textbf{Wrarning}

Do not invent CHUNK\_IDs. Only use CHUNK\_IDs that appear in the provided evidence items.

Do not cite the subsection title itself. Cite only evidence-based claims.

Do not include citations in any format other than [CHUNK\_ID].

Output ONLY the final paragraph text.

\end{tcolorbox}


\end{document}